\documentclass[a4paper,12pt]{article}
\pdfoutput=1
\usepackage{graphicx,subfigure,amsmath,amssymb}
\allowdisplaybreaks[4]
\usepackage{appendix}
\usepackage{color}
\usepackage{cite}

\newlength{\dinwidth}
\newlength{\dinmargin}
 \def\tb{\textcolor{black}}
\begin{document}
\title{\bf Study of $b\to c$ induced $\bar{B}^* \to V \ell \bar{\nu}_\ell$  decays }
\author{Qin Chang$^{a}$, Xiao-Lin Wang$^{a}$, Jie Zhu$^{b}$   and Xiao-Nan Li$^{b}$ \\
{ $^a$\small Institute of Particle and Nuclear Physics, Henan Normal University, Henan 453007, China}\\
{ $^b$\small School of physics and electronic engineering, Anyang Normal University, Henan 455000, China}} 
\date{}
 \maketitle

\begin{abstract}
In this paper, we investigate the tree-dominated $\bar{B}^*_{u,d,s,c} \to V \ell^- \bar{\nu}_\ell$ ($V=D^*_{u,d}\,,D^*_s\,,J/\psi$ and $\ell=e\,,\mu\,,\tau$)  decays in the  Standard Model with the relevant form factors obtained in the light-front quark model. These decays involve much more helicity states relative to the corresponding $\bar{B}^* \to P \ell^- \bar{\nu}_\ell$ and $\bar{B} \to V \ell^- \bar{\nu}_\ell$ decays, and moreover,  the contribution of  longitudinal polarization mode ($V$ meson) is relatively small, $\sim 30\%$, compared with the corresponding $B$ meson decays. We have also computed the branching fraction,  lepton spin asymmetry, forward-backward asymmetry  and ratio $R_V^{\ast(L)}\equiv \frac{\mathcal{B}(\bar{B}^*\to V\tau^- \bar{\nu}_\tau)}{\mathcal{B}(\bar{B}^*\to V \ell^{\prime-} \bar{\nu}_{\ell^{\prime}})}$  ($\ell'=e\,,\mu$).  Numerically,  the branching fractions of $\bar{B}^* \to  V \ell^{\prime-} \bar{\nu}_{\ell^{\prime}}$ decays are at the level of ${\cal O}(10^{-7})$, and are hopeful  to be observed by LHC and Belle-II experiments. \tb{The ratios $R_{D^*\,,D^*_s\,,J/\psi}^{\ast(L)}$ have relatively small theoretical uncertainties and are close to each other, $R^{*(L)}_{D^*}\simeq R^{*(L)}_{D^*_s}\simeq R^{*(L)}_{J/\psi} \simeq [0.26,0.27]~([0.27,0.29])$, which are a bit different from the predictions in some previous works. The future measurements are expected to make tests on these predictions.  }

\end{abstract}
\newpage

\section{Introduction}
In the past years, \tb{ a large amount of $B\bar{B}$ events have been accumulated by Babar, Belle, Tevatron and LHCb experiments, and most of $B$-meson decays having branching fractions $\gtrsim$ $\mathcal{O}(10^{-7})$ have been measured~\cite{Amhis:2016xyh}. Moreover,  some deviations between the standard model~(SM) predictions and the experimental data have been observed, for instance, the angular observable $P_5'$ of $B\to K^*\mu^+\mu^-$ decay with $2.6\sigma$ discrepancy~\cite{DescotesGenon:2012zf,Aaij:2015oid,Khachatryan:2015isa,Aaboud:2018krd,Wehle:2016yoi}, the differential branching fraction of $B_s\to \phi\mu^+\mu^-$  decay with $3.3\sigma$ discrepancy~\cite{Aaij:2015esa,Aaij:2013aln}, the well-known ``$\pi K$ CP puzzle''~\cite{Buras:2003dj,Buras:2004ub}, and so on.  Besides the flavor-changing-neutral-current precesses mentioned above,} the $B$-meson semileptonic decays induced by  $b \to c \ell \bar{\nu}_{\ell}$ transition also play an important role in testing the  SM and probing the  hints of possible new physics (NP). For instance, the well-known ``$R_{D^*}$ anomaly'' reported by BaBar~\cite{Lees:2012xj,Lees:2013uzd}, Belle~\cite{Huschle:2015rga,Sato:2016svk,Abdesselam:2016xqt} and LHCb~\cite{Aaij:2015yra,Aaij:2017uff} collaborations exhibits a significant deviation between the SM prediction and \tb{experimental data~\cite{Amhis:2016xyh,Jaiswal:2017rve,Jaiswal:2020wer}. Many studies have been done within the model-independent frameworks~\cite{Bhattacharya:2014wla,Bhattacharya:2015ida,Chen:2005gr,Bhattacharya:2016zcw,Bhattacharya:2018kig,Alok:2017qsi,Feruglio:2018fxo,Azatov:2018knx}, as well as in some specific NP models, for instance Refs.~\cite{Freytsis:2015qca,Li:2016vvp,Hiller:2016kry,Blanke:2018sro,Kim:2015zla,Crivellin:2015hha,He:2017bft,Hu:2020yvs,Wang:2019trs,Yan:2019hpm,Hu:2018veh,Li:2018rax,Altmannshofer:2017poe,Cheung:2020sbq,Celis:2012dk,Celis:2016azn,Li:2017jjs,Zhu:2016xdg,Wei:2018vmk,Hu:2018lmk}.  One can refer to Refs.~\cite{Li:2018lxi,Bifani:2018zmi} for recent reviews.
}

The spin-triplet vector $B^*_q$ meson with quantum number of $n^{2s+1}L_J=1^3S_1$ and $J^P=1^-$~\cite{Isgur:1991wq,Godfrey:1986wj,Eichten:1993ub,Ebert:1997nk} has the same flavor components as  the spin-singlet pseudoscalar $B_q$~($q=u,d,s$ and $c$)  \tb{meson, and}  can also decay through the $b \to c \ell \bar{\nu}_{\ell}$ transition at quark-level, therefore its $b\to c$ induced semileptonic decays  can play a similar role as $B$ meson decays for  testing the SM and probing possible hints of NP.

\tb{ The $B^*_q$ meson is unstable particle, it cannot decay via strong interaction due to that $m_{B^*_q}-m_{B_q}$$\lesssim50$ MeV$<$$m_{\pi}$~\cite{Tanabashi:2018oca}; $B^*_q$ meson decay is dominated by the radiative process~\cite{Tanabashi:2018oca},  $B^*_q\to B_q \gamma$; the weak decay modes via the bottom-changing transition~(for instance, the $b\to c$  induced semileptonic  $B^*_q$ decays considered in this work) are generally very rare, and their branching fractions are expected to be very small  within the SM.} Until now, there is no experimental information and few theoretical works concentrating on the $B^*_q$ weak decays. Fortunately, thanks to the high luminosity and large production cross section at the running LHC and SuperKEKB/Belle-II experiments, a huge amount of the ${B^*_q}$ meson data samples would be accumulated. \tb{At Belle-II experiment, the  $B^*$ and  $B^*_{s}$ mesons are  produced mainly via $\Upsilon(5S)$ decays. With the target annual integrated luminosity, $\sim$ 13 $ab^{-1}$~\cite{Abe:2010gxa}, and the cross section of $\Upsilon(5S)$ production in $e^+e^-$ collisions, $\sigma(e^+e^-\to\Upsilon(5S))=(0.301\pm0.002\pm0.039)\,{\rm nb}$~\cite{Huang:2006mf},  it is expected that about $4\times10^9$ $\Upsilon(5S)$ samples could be produced per year by Belle-II. Further considering that $\Upsilon(5S)$ meson mainly decays  to final states with a pair of $B^{(*)}_{(s)}$ mesons and using the branching fractions of $\Upsilon(5S)$ decays given by PDG~\cite{Tanabashi:2018oca}, it can be estimated that about $N(B^{*}+\bar{B}^{*})/{\rm year}\sim4\times10^9$ and $N(B^{*}_s+\bar{B}^{*}_s)/{\rm year}\sim2\times10^9$ samples can be accumulated by Belle-II per year. Unfortunately, the $B^{*}_c$ meson and its decays are out of the scope of  Belle-II experiment. In addition, a lot of $B^*_q$ samples can also  be produced via $pp$ collision and be accumulated in the future by LHC }with high collision energy, high luminosity and  rather large production cross section~\cite{Aaij:2010gn,Bediaga:2012py,Aaij:2014jba}, \tb{and some $B^*_q$ weak decays are hopeful to be observed, such as the leptonic $B_s^*\to\ell^+\ell^-$ decay with branching fraction $\sim {\cal O}(10^{-11})$~\cite{Grinstein:2015aua}. }

Encouraged by the abundant $B^*_q$ data samples at future heavy-flavor experiments, some interesting theoretical studies for the $B^*_q$ weak decays have been made within the SM, for instance,  the pure leptonic $\bar{B}_s^*\to \ell^+\ell^-$ and $\bar{B}_{u,c}^*\to \ell^- \bar{\nu}_\ell$  decays~\cite{Grinstein:2015aua}, the impact of  $\bar{B}_{s,d}^*\to \mu^+\mu^-$ on  $\bar{B}_{s,d}\to \mu^+\mu^-$ decays~\cite{Xu:2015eev},  the studies of the semileptonic $B^*_c$ decays within the QCD sum  rules~\cite{Wang:2012hu,Zeynali:2014wya,Bashiry:2014qia}, the semileptonic $B^*_{u,d,c,s}\to (P,V) \ell^- \bar{\nu}_{\ell}$ with $P=D,D_s,\eta_c,\,V=D^*,D^*_s,J/\psi$ decays within the Bethe-Salpeter (BS) method~\cite{Wang:2018dtb} and a approach under the assumption of  heavy quark symmetry (HQS)~\cite{Dai:2018vzz}, $\bar{B}^* \to P \ell^- \bar{\nu}_{\ell}$ with $P=D,D_s,\pi,K$~\cite{Chang:2016cdi} and the nonleptonic $\bar{B}^{*0}_{d,s}\to D_{d,s}^+M^-$  ($M=\pi\,,K\,,\rho$ and $K^*$)~\cite{Chang:2015jla,Chang:2015ead}, $\bar{B}_{d,s}^* \to D_{d,s} V$~\cite{Chang:2016eto}, $B^*_c \to B_{u,d,s}V, B_{u,d,s}P$~\cite{Sun:2017lup}, $B^*_c \to \eta_{c} V$~\cite{Chang:2018mva}, $B^* \to \bar{D}D$~\cite{Sun:2017xed} and $B^{\ast}_{c} \to \psi(1S,2S)P,\,\eta_c(1S,2S)P$~\cite{Sun:2017lla} decays. Moreover, the NP effects on the semileptonic $\bar{B}^* \to P \ell^- \bar{\nu}_{\ell}$ with $P=D,D_s,\pi,K$ decays have been investigated in a model-independent scheme~\cite{Chang:2018sud} and the vector leptoquark model~\cite{Zhang:2019hth}. In this paper, we pay our attention to the CKM-favored and tree-dominated semileptonic $\bar{B}^{\ast}_{u,d,s,c} \to V \ell \bar{\nu}_\ell$ $(V=D^{\ast}_{u,d},D^{\ast}_s,J/\psi)$ weak decays, which are generally much more complicated than the corresponding $B$ decay modes because they involve much more allowed helicity states.

Our paper is organized as follows. In section 2, the helicity amplitudes and observables of $\bar{B}^*\to V \ell \bar{\nu}_{\ell}$ decays are calculated. Section 3 is devoted to the numerical results and discussions, and  the $\bar{B}^*\to V$ transition form factors obtained within the covariant light-front quark mode are used in the computation. Finally, we give our summary in section 4.

\section{Theoretical framework and results}
\subsection{Effective Lagrangian and amplitude}
In the SM, $\bar{B}^{\ast}_{u,d,s,c} \to V \ell \bar{\nu}_\ell$ $(V=D^{\ast}_{u,d},D^{\ast}_s,J/\psi)$ decays are induced by $b\to c  \ell \bar{\nu}_\ell$
transition at quark level via W-exchange, and can be described by the effective Lagrangian
\begin{eqnarray}\label{eq:Leffall}
\mathcal{L}_{\rm eff}=-2\sqrt{2}
G_FV_{cb}\bar{c}_L\gamma^{\mu}b_{L}\bar{\ell}_L\gamma_{\mu}\nu_L
+\text{h.c.}\,,
\end{eqnarray}
at low energy scale $\mu={\cal O}(m_b)$, where $G_F$ is the Fermi coupling constant and $V_{cb}$ denotes the CKM matrix element. Using Eq.~\eqref{eq:Leffall}, the amplitude  of  $\bar{B}^* \to V \ell \bar{\nu}_\ell$ decay  can be written as the product of  hadronic matrix element and leptonic current. Then, in terms of leptonic ($L_{\mu\nu}$) and hadronic ($H^{\mu\nu}$) tensors built from the respective products of the leptonic and hadronic currents, the square amplitude can be expressed  as
\begin{eqnarray}\label{eq:M2}
|{\cal M}(\bar{B}^{\ast} \to V \ell^- \bar{\nu}_\ell)|^2=|\langle V \ell^- \bar{\nu}_\ell|\mathcal{L}_{\rm eff}|\bar{B}^{\ast}\rangle|^2=\frac{G_F^2|V_{cb}|^2}{2}L_{\mu\nu} H^{\mu\nu}\,.
\end{eqnarray}
Inserting the completeness relation of  the polarization vector of  virtual $W^*$ boson,
\begin{eqnarray}
\sum_{m,n}\bar{\epsilon}_{\mu}(m) \bar{\epsilon}_{\nu}^*(n)g_{mn}=g_{\mu\nu}\,,
\end{eqnarray}
the product of $L_{\mu\nu}$ and $H^{\mu\nu}$ can be rewritten as
\begin{eqnarray}\label{eq:M2LI}
L_{\mu\nu}H^{\mu\nu}=\sum_{m,m^{\prime},n,n^{\prime}} L(m,n)H(m^{\prime},n^{\prime})g_{mm^{\prime}}g_{nn^{\prime}}\,,
\end{eqnarray}
where $L(m,n)\equiv  L^{\mu\nu}\bar{\epsilon}_{\mu}(m)\bar{\epsilon}^*_{\nu}(n)$ and $H(m,n)\equiv  H^{\mu\nu}\bar{\epsilon}^*_{\mu}(m)\bar{\epsilon}_{\nu}(n)$ are Lorentz invariant and therefore can be evaluated in different reference frames. In our following evaluation, $H(m,n)$ and $L(m,n)$ will be calculated  in the $B^*$-meson rest frame and the  $\ell-\bar{\nu}_\ell$ center-of-mass frame, respectively.

\subsection{Kinematics}
In the rest frame of  $B^*$ meson, assuming the final state V-meson moving along with positive $z$-direction, the momenta of  $B^*$, $V$ and $W^*$ could be written as
\begin{eqnarray}
 p_{B^*}^{\mu}=(m_{B^*},0,0,0)\,,\quad  p_{V}^{\mu}=(E_{V},0,0,|\vec{p}|)\,,\quad  q^{\mu}=(q^0,0,0,-|\vec{p}|)\,,
\end{eqnarray}
respectively, where $q^0=(m_{B^*}^2-m_{V}^2+q^2)/2m_{B^*}$ and $|\vec{p}|=\lambda^{1/2}(m_{B^*}^2,m_{V}^2,q^2)/2m_{B^*}$, with $\lambda(a,b,c)\equiv a^2+b^2+c^2-2(ab+bc+ca)$ and $q^2=(p_{B^*}-p_{V})^2$ being the momentum transfer squared, are the energy and momentum of  virtual $W^*$. The polarization vectors of the  initial $B^*$-meson and daughter $V$-meson, $\epsilon_1^{\mu}(0,\pm)$ and $\epsilon_2^{\mu}(0,\pm)$, can be written as
 \begin{eqnarray}\label{eq:polBstar}
&&\epsilon_1^{\mu}(0)=(0,0,0,1)\,,\quad  \epsilon_1^{\mu}(\pm)=\frac{1}{\sqrt{2}}(0,\mp1,-i,0)\,;\\\label{eq:polV}
&&\epsilon_2^{\mu}(0)=\frac{1}{m_{V}}(|\vec{p}|,0,0,E_{V})\,,\quad  \epsilon_2^{\mu}(\pm)=\frac{1}{\sqrt{2}}(0,\mp1,-i,0)\,,
\end{eqnarray}
respectively. For the four polarization vectors of virtual $W^*$, $\bar{\epsilon}^{\mu}(t,0,\pm)$, one can conveniently choose~\cite{Korner:1987kd,Korner:1989qb}
 \begin{eqnarray}\label{eq:polW}
\bar{\epsilon}^{\mu}(t)=\frac{1}{\sqrt{q^2}}(q^0,0,0,-|\vec{p}|)\,,\quad \bar{\epsilon}^{\mu}(0)=\frac{1}{\sqrt{q^2}}(|\vec{p}|,0,0,-q^0)\,,\quad  \bar{\epsilon}^{\mu}(\pm)=\frac{1}{\sqrt{2}}(0,\pm1,-i,0)\,,
\end{eqnarray}
in which, $\lambda_{W^*}=t$ has to be understood as $\lambda_{W^*}=0$ and $J=0$.

Turning to the $\ell-\bar{\nu}_\ell$ center-of-mass frame, the four-momenta of lepton and antineutrino  are given as
 \begin{eqnarray}
 p_\ell^{\mu}=(E_{\ell}, |\vec{p}_{\ell}|\sin\theta,0,|\vec{p}_{\ell}|\cos\theta)\,,\quad  p_{\nu_\ell}^{\mu}=(|\vec{p}_{\ell}|, -|\vec{p}_{\ell}|\sin\theta,0,-|\vec{p}_{\ell}|\cos\theta)\,,
 \end{eqnarray}
where  $E_{\ell}=(q^2+m_{\ell}^2)/2\sqrt{q^2}$, $|\vec{p}_{\ell}|=(q^2-m_{\ell}^2)/2\sqrt{q^2}$, and $\theta$ is the angle between  $V$ and ${\ell}$ three-momenta. In this frame, the polarization vectors $\bar{\epsilon}^{\mu}(\lambda_{W^*})$  have the form
 \begin{eqnarray}\label{eq:polWl}
\bar{\epsilon}^{\mu}(t)=(1,0,0,0)\,,\quad \bar{\epsilon}^{\mu}(0)=(0,0,0,1)\,, \quad \bar{\epsilon}^{\mu}(\pm)=\frac{1}{\sqrt{2}}(0,\mp1,-i,0)\,.
\end{eqnarray}

\subsection{Hadronic helicity amplitudes}
For hadronic part, one has to calculate  the hadronic helicity amplitudes $H_{\lambda_{W^{\ast}}\lambda_{B^{\ast}}\lambda_{V} }$ of  $\bar{B}^{\ast} \to V \ell^- \bar{\nu}_\ell$ decay defined by
 \begin{eqnarray}\label{eq:hll}
 H_{ \lambda_{W^*}\lambda_{B^*}\lambda_{V}}(q^2)=\langle V(p_{V},\,\lambda_{V})|\bar{c}\gamma_{\mu}(1-\gamma_5)b|\bar{B}^*(p_{B^*},\,\lambda_{B^*})\rangle\bar{\epsilon}^{ *\mu}(\lambda_{W^*})\,,
\end{eqnarray}
which describes the decay of three helicity states of $B^*$ meson into the three helicity states of daughter $V$ meson and the four helicity states of  virtual $W^*$. For the $B^*\to V$ transition, the matrix elements  $\langle V(p_{V},\,\lambda_{V})|\bar{c}\gamma_{\mu} (1-\gamma_{5} )b|\bar{B}^*(p_{B^*},\,\lambda_{B^*})\rangle $ can be factorized in terms of ten form factors $V_{1,2,3,4,5,6}(q^2)$ and $A_{1,2,3,4}(q^2)$ as~\cite{Wang:2007ys,Shen:2008zzb}
\begin{eqnarray}
\langle V(\epsilon_2, p_{V})|\bar{c}\gamma_{\mu} b|\bar{B}^*(\epsilon_1, p_{B^*})\rangle
&=&(\epsilon_1 \cdot\epsilon_2^{*})\left[-P_{\mu}\,V_1(q^2)+q_{\mu}\,V_2(q^2)\right]\nonumber\\
&&+\frac{(\epsilon_1 \cdot q)(\epsilon_2^{*}\cdot q)}{m_{B^*}^2-m_{V}^2}\left[P_{\mu}\,V_3(q^2)-q_{\mu}\,V_4(q^2)\right]\nonumber\\
&&-(\epsilon_1 \cdot q)\,\epsilon_{2\mu}^* \,V_5(q^2)+(\epsilon_2^* \cdot q)\,\epsilon_{1\mu}\,V_6(q^2)\,,\\
\langle V(\epsilon_2, p_V)|\bar{c} \gamma_{5} \gamma_{\mu} b|\bar{B}^*(\epsilon_1, p_{B^*})\rangle
&=&-i\varepsilon_{\mu\nu\alpha\beta}\epsilon_1^{\alpha}\epsilon_2^{*\beta}\left[
P^{\nu}\,A_1(q^2)-q^{\nu}\,A_2(q^2)\right]\nonumber\\
&&-\frac{i\epsilon_2^{*}\cdot q}{m_{B^*}^2-m_{V}^2}\varepsilon_{\mu\nu\alpha\beta}\epsilon_1^{\nu}P^{\alpha}q^{\beta}\,A_3(q^2)\nonumber\\
&&+\frac{i\epsilon_1\cdot q}{m_{B^*}^2-m_{V}^2}\varepsilon_{\mu\nu\alpha\beta}\epsilon_2^{*\nu}P^{\alpha}q^{\beta}A_4(q^2)
\end{eqnarray}
with the sign convention $\epsilon_{0123}=-1$.

Then, by contracting these hadronic matrix elements with the polarization vector of virtual $W^*$ boson, we can finally obtain the non-vanishing hadronic helicity amplitudes,  $H_{ \lambda_{W^*}\lambda_{B^*}\lambda_{V}}$, given as
\begin{eqnarray}
H_{0++}(q^2)&=&-\frac{m_{B^*}^2-m_{V}^2} {\sqrt{q^2}}A_1(q^2)+\sqrt{q^2}A_2(q^2)+\frac{2m_{B^*}|\vec{p}|}{\sqrt{q^2}}V_1(q^2)\,,\\
H_{t++}(q^2)&=&-\frac{2m_{B^*}|\vec{p}|}{\sqrt{q^2}}A_1(q^2)+\frac{m_{B^*}^2-m_{V}^2}{\sqrt{q^2}}V_1(q^2)-\sqrt{q^2}V_2(q^2)\,,\\
H_{-+0}(q^2)&=&-\frac{m_{B^*}^2+3m_{V}^2-q^2}{2m_{V}}A_1(q^2)+\frac{(m_{B^*}^2-m_{V}^2-q^2)}{2m_{V}}A_2(q^2)\nonumber\\
                &&-\frac{2m_{B^*}^2|\vec{p}|^2}{m_{V}(m_{B^*}^2-m_{V}^2)}A_3(q^2)
                -\frac{m_{B^*}|\vec{p}|}{m_{V}}V_6(q^2)\,,\\
H_{0--}(q^2)&=&\frac{m_{B^*}^2-m_{V}^2}{\sqrt{q^2}}A_1(q^2)-\sqrt{q^2}A_2(q^2)+\frac{2m_{B^*}|\vec{p}|}{\sqrt{q^2}}V_1(q^2)\,,\\
H_{t--}(q^2) &=&\frac{2m_{B^*}|\vec{p}|}{\sqrt{q^2}}A_1(q^2)+\frac{m_{B^*}^2-m_{V}^2}{\sqrt{q^2}}V_1(q^2)
            -\sqrt{q^2}V_2(q^2)\,,\\
H_{+-0}(q^2)&=&\frac{m_{B^*}^2+3m_{V}^2-q^2}{2m_{V}}A_1(q^2)-\frac{(m_{B^*}^2-m_{V}^2-q^2)
            }{2m_{V}}A_2(q^2)\nonumber\\
            &&+\frac{2m_{B^*}^2|\vec{p}|^2}{m_{V}(m_{B^*}^2-m_{V}^2)}A_3(q^2)
            -\frac{m_{B^*}|\vec{p}|}{m_{V}}V_6(q^2)\,,\\
H_{+0+}(q^2)&=&\frac{3m_{B^*}^2+m_{V}^2-q^2}{2m_{B^*}}A_1(q^2)
            -\frac{(m_{B^*}^2-m_{V}^2+q^2)}{2m_{B^*}}A_2(q^2)\nonumber\\
            &&+\frac{2m_{B^*}|\vec{p}|^2}{m_{B^*}^2-m_{V}^2}A_4(q^2)-|\vec{p}|V_5(q^2)\,,\\
H_{-0-}(q^2)&=&-\frac{3m_{B^*}^2+m_{V}^2-q^2}{2m_{B^*}}A_1(q^2)
            +\frac{(m_{B^*}^2-m_{V}^2+q^2)}{2m_{B^*}}A_2(q^2)\nonumber\\
            &&-\frac{2m_{B^*}|\vec{p}|^2}{m_{B^*}^2-m_{V}^2}A_4(q^2)-|\vec{p}|V_5(q^2)\,,\\
H_{000}(q^2)&=&\frac{|\vec{p}|(m_{B^*}^2+m_{V}^2-q^2)}{\sqrt{q^2}m_{V}}V_1(q^2)
            +\frac{2m_{B^*}^2|\vec{p}|^3}{\sqrt{q^2}m_{V}(m_{B^*}^2-m_{V}^2)}V_3(q^2)\nonumber\\
            &&-\frac{|\vec{p}|(m_{B^*}^2-m_{V}^2-q^2)}{2\sqrt{q^2}m_{V}}V_5(q^2)
            +\frac{|\vec{p}|(m_{B^*}^2-m_{V}^2+q^2)}{2\sqrt{q^2}m_{V}}V_6(q^2)\,,\\
H_{t00}(q^2) &=&\frac{(m_{B^*}^2-m_{V}^2)(m_{B^*}^2+m_{V}^2-q^2)}{2\sqrt{q^2}m_{B^*}m_{V}}V_1(q^2)
            -\frac{\sqrt{q^2}(m_{B^*}^2+m_{V}^2-q^2)}{2m_{B^*}m_{V}}V_2(q^2)\nonumber\\
            &&+\frac{m_{B^*}|\vec{p}|^2}{\sqrt{q^2}m_{V}}V_3(q^2)
            -\frac{m_{B^*}|\vec{p}|^2\sqrt{q^2}}{m_{V}(m_{B^*}^2-m_{V}^2)}V_4(q^2)
            -\frac{m_{B^*}|\vec{p}|^2}{\sqrt{q^2}m_{V}}V_5(q^2)\nonumber\\
           &&+\frac{m_{B^*}|\vec{p}|^2}{\sqrt{q^2}m_{V}}V_6(q^2)\,.
\end{eqnarray}
Obviously, only the amplitudes with $\lambda_{B^*}=\lambda_{V}-\lambda_{W^*}$ survive due to the helicity conservation.

\subsection{Helicity  amplitudes and observables}
For the leptonic part, the leptonic tensor could be expanded in terms of a complete set of Wigner's $d^J$-functions, which has been widely used in the study of hadron semileptonic~\cite{Fajfer:2012vx,Korner:1987kd,Kadeer:2005aq}. As a result, $L_{\mu\nu}H^{\mu\nu}$ can be reduced to a very compact form
\begin{eqnarray}\label{eq:LH}
L_{\mu\nu}H^{\mu\nu}&=&\frac{1}{8}\sum_{ \substack{\lambda_{\ell},\lambda_{\bar{\nu}_{\ell}}, J' ,J \\    \lambda_{W^*}, \lambda_{W^*}'} }
                    (-1)^{J+J'}|h_{\lambda_{\ell},\,\lambda_{\bar{\nu}_{\ell}}}|^2
                    \delta_{\lambda_{B^*},\,\lambda_{V}-\lambda_{W^*}}\,\delta_{\lambda_{B^*},\,\lambda_{V}-\lambda'_{W^*}}\nonumber\\
                    &&\times\,d_{\lambda_{W^*},\,\lambda_{\ell}-\frac{1}{2}}^J\,d_{\lambda'_{W^*},\,\lambda_{\ell}-\frac{1}{2}}^{J'}\,
                    H_{\lambda_{W^*}\lambda_{B^*}\lambda_{V}}\,H_{\lambda'_{W^*}\lambda_{B^*}\lambda_{V}}
\end{eqnarray}
where $J$ and $J'$ run over 1 and 0, $\lambda^{(')}_{W^{\ast}}$ and $\lambda_{\ell}$ run over their components. For the standard expression of $d^J$ function, we take their value from  PDG~\cite{Tanabashi:2018oca}. The leptonic helicity amplitude $h_{\lambda_{\ell},\lambda_{\bar{\nu}_{\ell}}}$ in
Eq.~\eqref{eq:LH} defined as
\begin{eqnarray}
h_{\lambda_{\ell},\lambda_{\bar{\nu}_{\ell}}}=\bar{u}_{\ell}(\lambda_{\ell})\gamma^{\mu}(1-\gamma_5)\nu_{\bar{\nu}}(\frac{1}{2})\bar{\epsilon}_{\mu}(\lambda_{W^*})\,,
\end{eqnarray}
Taking the exact forms of spinors and $W^{\ast}$ polarization vectors given in Eq.~\eqref{eq:polWl}, we obtain
\begin{eqnarray}
|h_{-\frac{1}{2},\frac{1}{2}}|^2=8(q^2-m_\ell^2)\,,\qquad
|h_{\frac{1}{2},\frac{1}{2}}|^2=8\frac{m_\ell^2}{2q^2}(q^2-m_\ell^2)\,,
\end{eqnarray}
which are the same as the results obtained in semileptonic $B$ and hyperon decays~\cite{Fajfer:2012vx,Kadeer:2005aq}.

Using the amplitudes obtained above, we can then further evaluate the observables of $\bar{B}^* \to V \ell^- \bar{\nu}_\ell$ decays.
The double differential decay rate is  written as
\begin{eqnarray}
\frac{d^2\Gamma}{dq^2d\cos\theta}=\frac{G_F^2|V_{cb}|^2}{(2\pi)^3}\,\frac{|\vec{p}|}{8m_{B^*}^2}\,
            \frac{1}{3}(1-\frac{m_\ell^2}{q^2})L_{\mu\nu}H^{\mu\nu}\,,
\end{eqnarray}
where the factor $1/3$ is caused by averaging over the spins of initial $\bar{B}^*$ meson.
The double differential decay rate with a given helicity state of lepton ($\lambda_{\ell}=\pm \frac{1}{2}$) is written as
\begin{eqnarray}
\label{eq:DdGml}
\frac{d^2\Gamma\left[\lambda_{\ell}=-1/2\right]}{dq^2d\cos\theta}&=&\frac{G_F^2|V_{cb}|^2|\vec{p}|}{256\pi^3m_{B^*}^2}\,
        \frac{1}{3}q^2(1-\frac{m_\ell^2}{q^2})^2\,\nonumber\\
        &&\times\Big[(1-\text{cos}\,\theta)^2(H_{+0+}^2+H_{+-0}^2)
        +(1+\text{cos}\,\theta)^2(H_{-0-}^2+H_{-+0}^2)\nonumber\\
        &&+2\text{sin}^2\,\theta(H_{0++}^2+H_{0--}^2+H_{000}^2)\Big],\\
\label{eq:DdGpl}
\frac{d^2\Gamma\left[\lambda_{\ell}=1/2\right]}{dq^2d\cos\theta}&=&\frac{G_F^2|V_{cb}|^2|\vec{p}|}{256\pi^3m_{B^*}^2}\,
        \frac{1}{3}q^2(1-\frac{m_\ell^2}{q^2})^2\,\frac{m_{\ell}^2}{q^2}\,\nonumber\\
        &&\Big[\text{sin}^2\,\theta(H_{+0+}^2+H_{+-0}^2+H_{-0-}^2+H_{-+0}^2)
        +2(H_{t++}-\text{cos}\,\theta\,H_{0++})^2\nonumber\\
         &&+2(H_{t--}-\text{cos}\,\theta\,H_{0--})^2
        +2(H_{t00}-\text{cos}\,\theta\,H_{000})^2\Big].
\end{eqnarray}

Integrating over $\text{cos}\,\theta$ and  summing over  the lepton helicity, we can obtain the differential decay rate  written as
\begin{eqnarray}\label{eq:DdG}
\frac{d\Gamma}{dq^2}&=&\frac{G_F^2|V_{cb}|^2|\vec{p}|}{96\pi^3m_{B^*}^2}\,
        \frac{1}{3}\,q^2\,(1-\frac{m_\ell^2}{q^2})^2\,\times
        \bigg[\frac{3m_{\ell}^2}{2q^2}(H_{t++}^2+H_{t--}^2+H_{t00}^2)\nonumber\\
        &&+(H_{+0+}^2+H_{+-0}^2+H_{-0-}^2+H_{-+0}^2+H_{000}^2+H_{0--}^2+H_{0++}^2)(1+\frac{m_{\ell}^2}{2q^2})\bigg]\,,
\end{eqnarray}
in which, the three non-diagonal interference terms in Eq.~\eqref{eq:DdGpl} vanish.  In addition,  paying attention to the polarization  states of  $V$ meson, one can obtain the  longitudinal differential decay width $d\Gamma^L/dq^2$ by picking out  $H^2_{t00}$, $H^2_{+-0}$, $H^2_{-+0}$ and $H^2_{000}$ terms in Eq.~\eqref{eq:DdG}.

 Using Eqs.~\eqref{eq:DdGml} and \eqref{eq:DdGpl} given above, we can also construct some useful observables as follows.
The $q^2$ dependent ratios is defined as
  \begin{eqnarray}\label{eq:dR}
  R^{*(L)}_{V}(q^2)\equiv \frac{d\Gamma^{(L)}(\bar{B}^*\to V \tau^-\bar{\nu}_\tau)/dq^2}{d\Gamma^{(L)}(\bar{B}^*\to V\ell^{\prime-}\bar{\nu}_{\ell^{\prime}})/dq^2}\,,
  \end{eqnarray}
where, $\ell^{\prime}$ denotes the light leptons $\mu$ and $e$ (in the following calculations, we take $m_{e,\mu}=0$). 
 The lepton spin asymmetry and forward-backward asymmetry are defined as
  \begin{eqnarray}
  \label{eq:ALamb}
A^{*V}_{\lambda}(q^2)&=&\frac{d\Gamma[\lambda_\ell=-1/2]/dq^2-d\Gamma[\lambda_\ell=1/2]/dq^2}
{d\Gamma[\lambda_\ell=-1/2]/dq^2+d\Gamma[\lambda_\ell=1/2]/dq^2}\,,
\end{eqnarray}
   and
   \begin{eqnarray}
  \label{eq:ATheta}
A^{*V}_{\theta}(q^2)&=&\frac{\int_{-1}^0d\cos\theta\,( d^2\Gamma/dq^2d\cos\theta)-\int_{0}^1d\cos\theta\,( d^2\Gamma/dq^2d\cos\theta)}{ d\Gamma/dq^2}\,,
\end{eqnarray}
respectively. These observables are independent of the CKM matrix elements, and the hadronic  uncertainties canceled to a large extent, therefore, they can be predicted with a rather high accuracy.

\section{Numerical results and discussions}
In our numerical calculation, for the well-known Fermi coupling constant $G_F$ and the masses of mesons and $\tau$, we take their central values given by PDG~\cite{Tanabashi:2018oca}. For  the CKM element, we take $|V_{cb}|=41.80^{+0.28}_{-0.60}\times 10^{-3} $ given by CKMFitter Group~\cite{Charles:2004jd}. In order to evaluate the branching fractions, the total decay widths (or lifetimes), $\Gamma_{\rm tot}(B^*_{u,d,s,c})$, are also essential inputs. However, there is no available experimental or theoretical information until now. While, due to the fact that the electromagnetic processes $B^* \to B \gamma$ dominates $B^*$ decays, we can take the approximation $\Gamma_{\rm tot}(B^*) \simeq \Gamma(B^* \to B \gamma)$.
In the light-front quark model~(LFQM), the decay width of $B^* \to B \gamma$ decay is given by~\cite{Choi:2007se}
\begin{eqnarray}
\Gamma(B^* \to B \gamma)&=&\frac{\alpha}{3}\,[e_1I(m_1,m_2,0)+e_2\,I(m_2,m_1,0)]^2\,\kappa^3_{\gamma}\,,\\
I(m_1\,, m_2\,, q^2)&=&\int^1_0\frac{dx}{8 \pi^3}\int d^2k_{\perp}\,
\frac{\psi(x\,,k'_{\perp})\,\psi(x\,,k_{\perp})}{x\,\tilde{M}_0\,\tilde{M}'_0}
\times\left\{\mathcal{A}+\frac{2}{\mathcal{M}_{0}}\,[k^2_{\perp}-\frac{(k_{\perp}\cdot q_{\perp})^2}{q^2_{\perp}}]\right\}\,,
\end{eqnarray}
where $\mathcal{A}=\bar{x}m_1+xm_2$ with $\bar{x}=1-x$, $\mathcal{M}_{0}=M_0+m_1+m_2$ with $M_0$ being the invariant  mass of bound-state, $\alpha$ is the fine-structure constant, $\kappa_{\gamma}=(m^2_{B^*}-m^2_{B})^2/2m_{B^*}$ is the kinematically allowed energy of the outgoing photon. The radial wavefunction~(WF) $\psi(x,k_\perp)$ of bound-state is responsible for describing the momentum distribution of the constituent quarks. In this paper, we shall use the Gaussian-type WF
\begin{align}
\label{eq:RWFs}
\psi(x,{k}_{\bot}) =4\frac{\pi^{\frac{3}{4}}}{\beta^{\frac{3}{2}}} \sqrt{ \frac{\partial k_z}{\partial x}}\exp\left[ -\frac{k_z^2+{k}_\bot^2}{2\beta^2}\right]\,,
\end{align}
 where $k_z$ is the relative momentum in $z$-direction and has the form $ k_z=(x-\frac{1}{2})M_0+\frac{m_2^2-m_1^2}{2 M_0}$.
One can refer to Ref.~\cite{Choi:2007se} for more details. Using  the constituent quark masses and the  Gaussian parameter $\beta$ given in Table \ref{tab:LFinput}, we obtain the numerical results for $\Gamma(B^{*}\to B \gamma)$ as follows,
\begin{eqnarray}
\label{eq:GtotBu}
\Gamma_{\rm{tot}}(B^{*+})&\simeq& \Gamma(B^{*+}\to B^+ \gamma)=(349\pm{18})\,{\rm eV},\\
 \label{eq:GtotBd}
\Gamma_{\rm{tot}}(B^{*0})&\simeq& \Gamma(B^{*0}\to B^0 \gamma)=(116\pm6)\,{\rm eV},\\
 \label{eq:GtotBs}
\Gamma_{\rm{tot}}(B^{*0}_s)&\simeq& \Gamma(B^{*0}_s\to B^0_s \gamma)=(84^{+11}_{-9})\,{\rm eV},\\
 \label{eq:GtotBc}
\Gamma_{\rm{tot}}(B^{*+}_c)&\simeq& \Gamma(B^{*+}_c\to B^0_c \gamma)=(49^{+28}_{-21})\,{\rm eV}.
\end{eqnarray}
These theoretical predictions are generally in agreement with the ones obtained in the previous work based on different theoretical models~\cite{Goity:2000dk,Ebert:2002xz,Zhu:1996qy,Aliev:1995wi,Colangelo:1993zq, Choi:2007se,Cheung:2014cka}.

\begin{table}[t]
\caption{The values of constituent quark masses and Gaussian parameters (in units of MeV) obtained by fitting to the data of decay constants~\cite{Verma:2011yw,Chang:2018zjq}, where $q=u\,,d$.}
\begin{center}
\begin{tabular}{l}
\hline\hline
$m_q=250$\,, $m_s=450$\,, $m_c=1400$\,, $m_b=4640\,$;\\\hline
$\beta_{b\bar{q}}=540.7\pm9.6$\,, $\beta_{b\bar{s}}=601.9\pm7.4$\,, $\beta_{b\bar{c}}=933.9\pm11.1$\,, ~~~~~~\text{for P-meson}\\\hline
$\beta_{c\bar{q}}=413.0\pm12.0$\,, $\beta_{c\bar{s}}=514.1\pm18.5$\,, $\beta_{c\bar{c}}=684.4\pm6.7$\,,\\
$\beta_{b\bar{q}}=504.4\pm14.2$\,, $\beta_{b\bar{s}}=556.4\pm10.1$\,, $\beta_{b\bar{c}}=863.4\pm32.8$\,, ~~~\text{for V-meson}\\\hline
\hline\hline
\end{tabular}
\end{center}
\label{tab:LFinput}
\end{table}
%

\begin{figure}[ht]
\caption{The $q^2$-dependences of form factors for $\bar{B}^* \to D^*$, $\bar{B}^*_s \to D^*_s$ and $\bar{B}^*_c \to J/\psi$ transitions.}
\begin{center}
\subfigure[]{\includegraphics[scale=0.45]{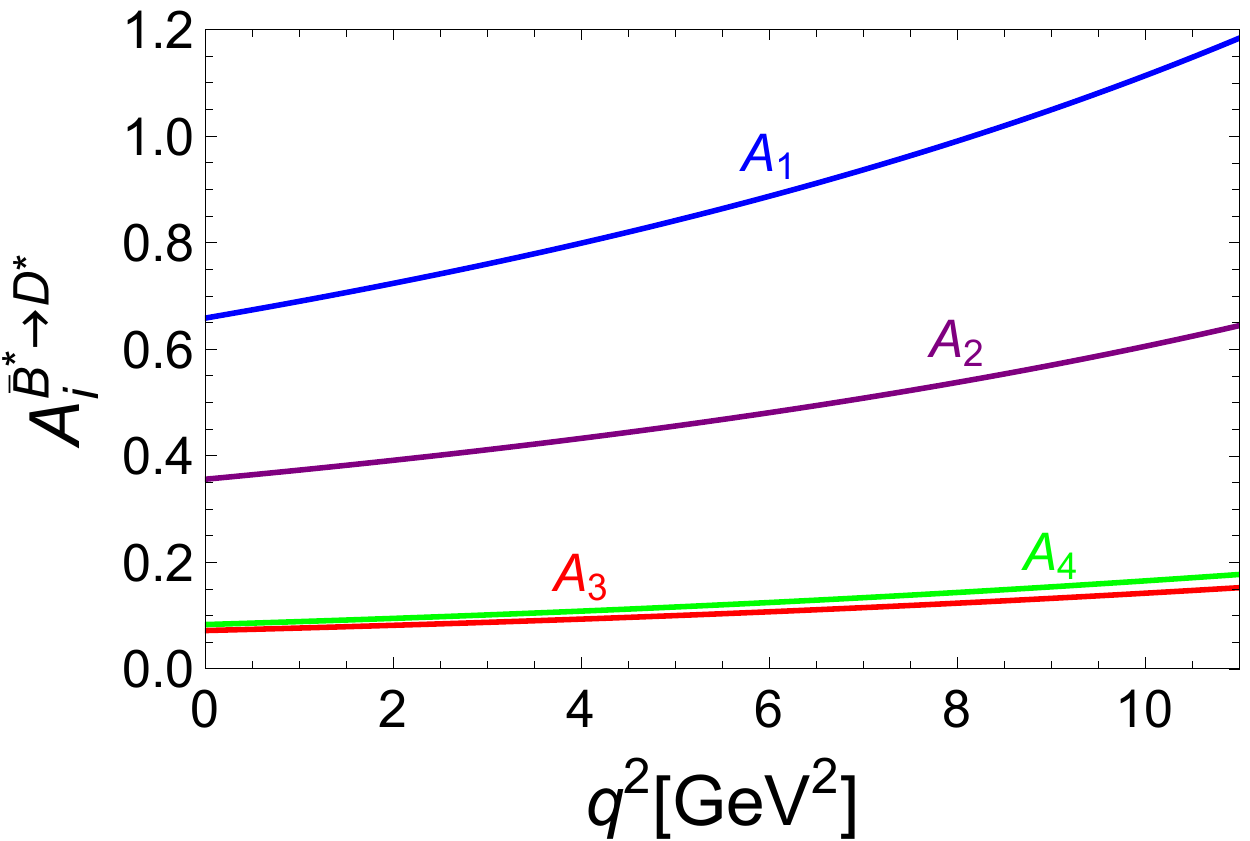}}\qquad
\subfigure[]{\includegraphics[scale=0.45]{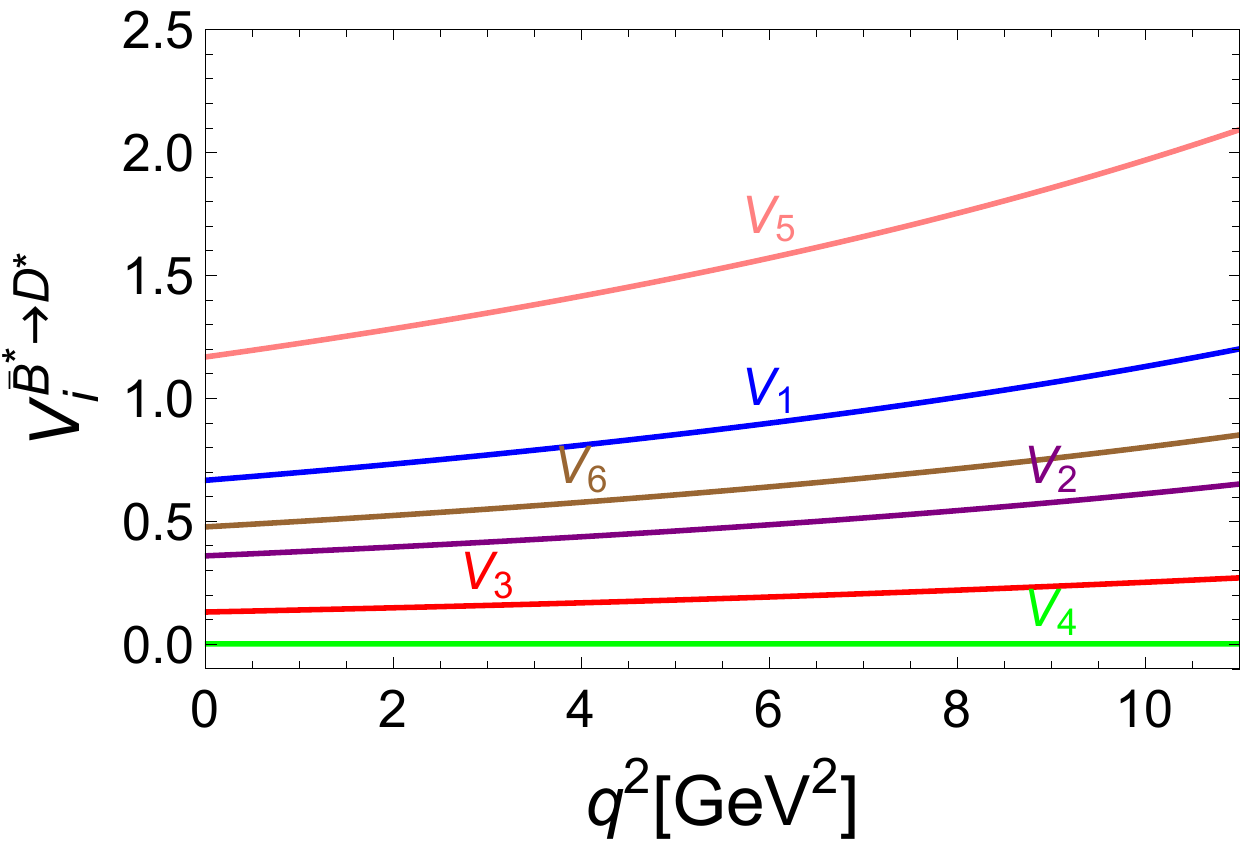}}\\
\subfigure[]{\includegraphics[scale=0.45]{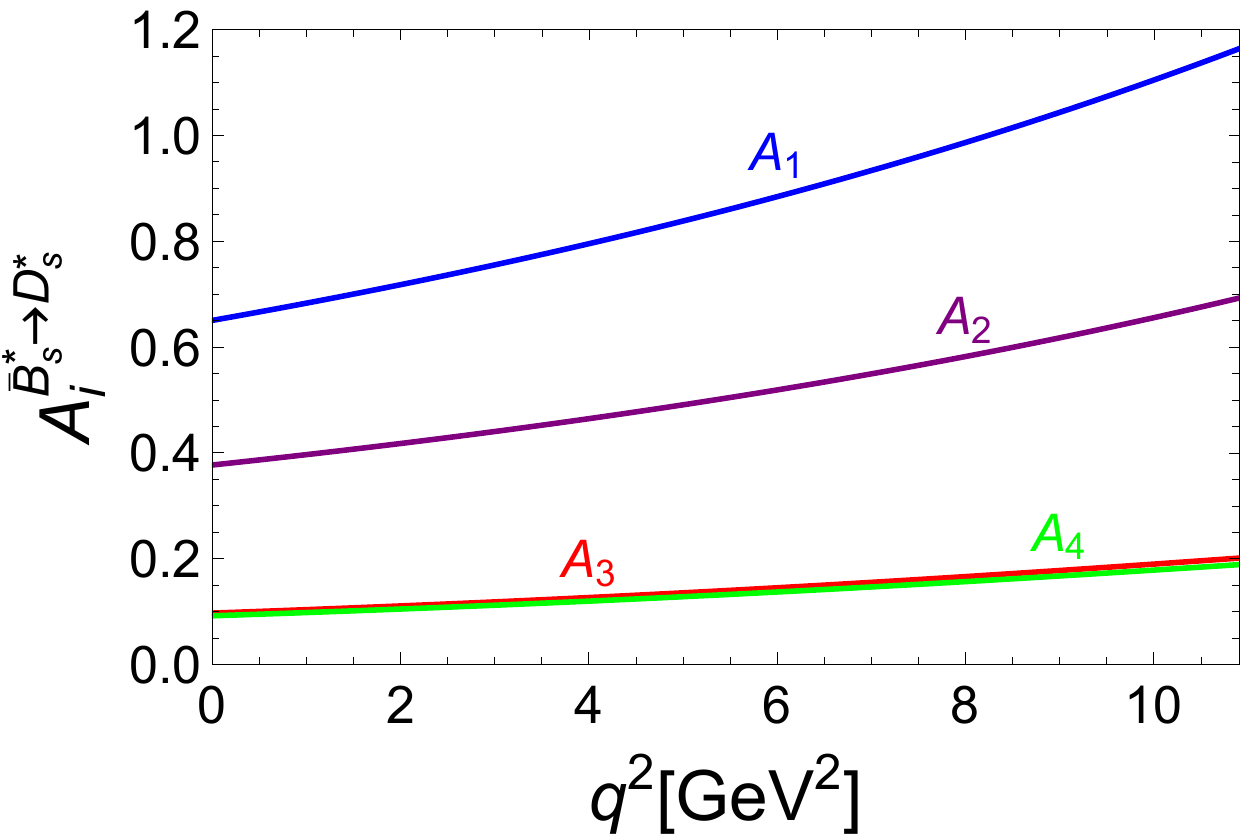}}\qquad
\subfigure[]{\includegraphics[scale=0.45]{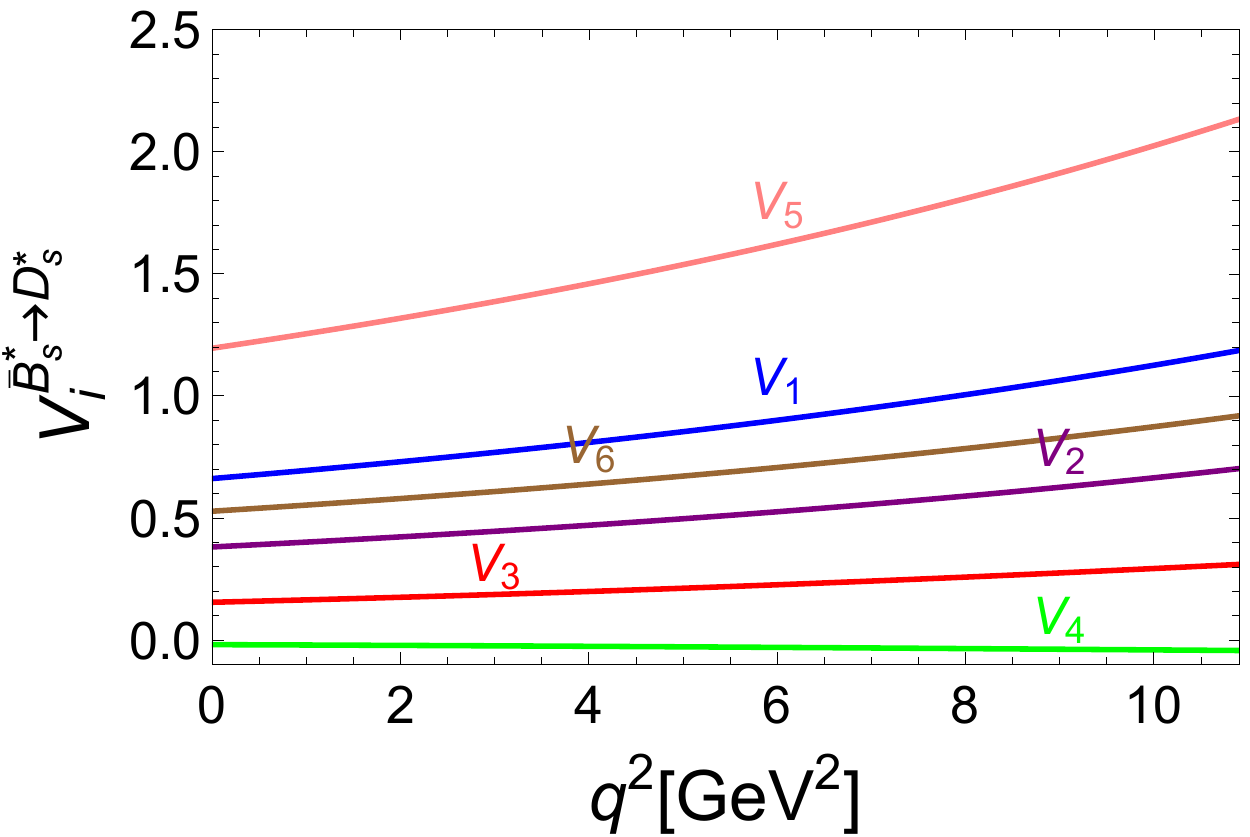}}\\
\subfigure[]{\includegraphics[scale=0.45]{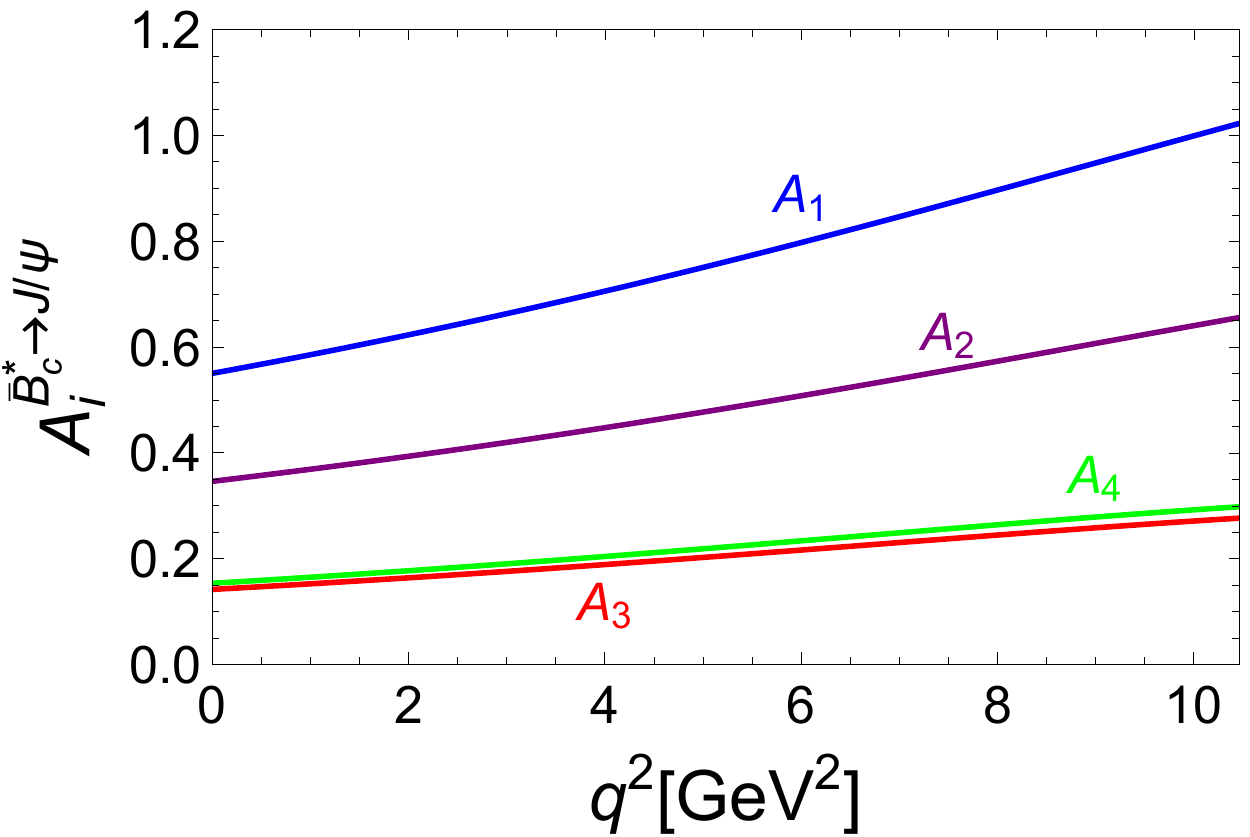}}\qquad
\subfigure[]{\includegraphics[scale=0.45]{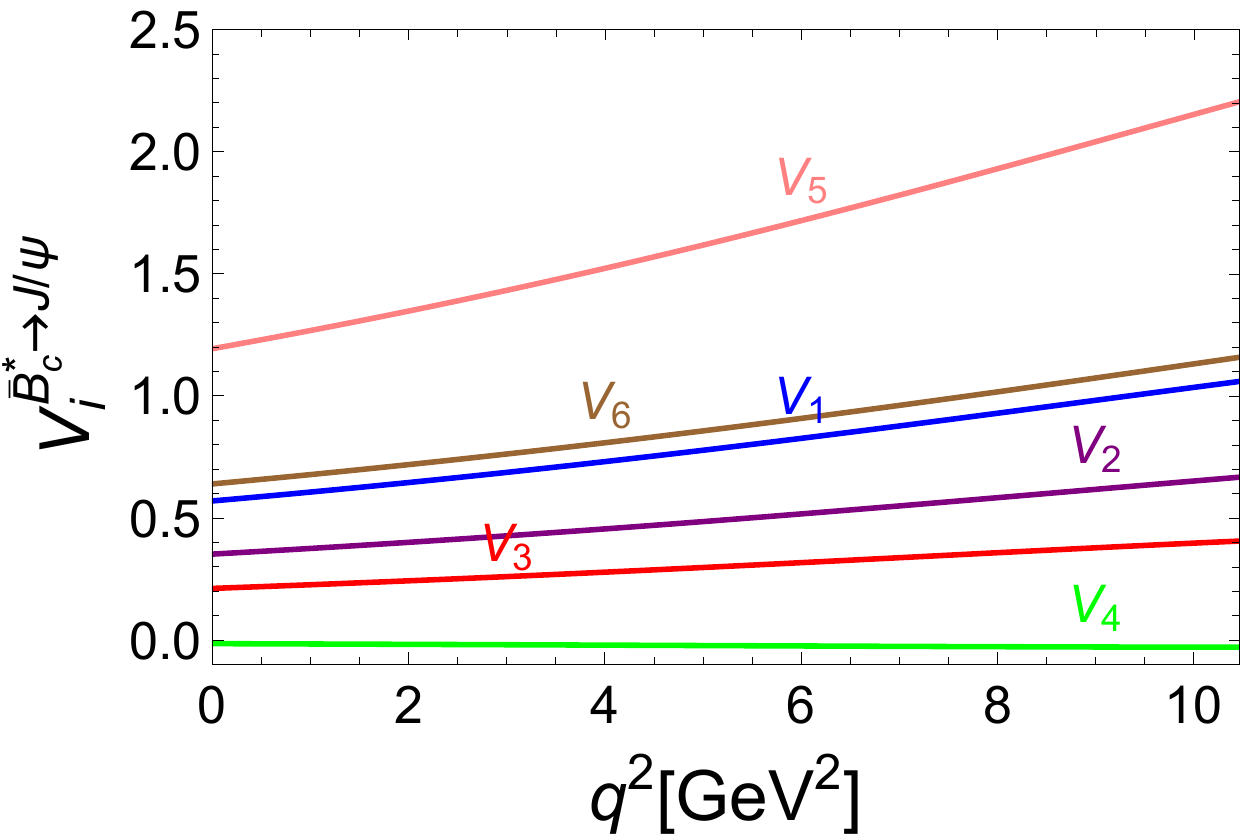}}
\end{center}
\label{fig:FFBstar2V}
\end{figure}

Besides the  inputs given above, the $B^{*}\to V$ transition form factors are also crucial inputs for evaluating observables, especially for the branching fraction. In this work, we adopt the covariant light-front quark model~(CLFQM)~\cite{Jaus:1999zv,Jaus:2002sv,Cheng:2003sm}  to evaluate their values. The theoretical formulas for the form factors of $V'\to V''$ have been given in our previous work~(see Eqs.~(39-48) in the appendix of Ref.~\cite{Chang:2019xtj}). These theoretical results are obtained within Drell-Yan-West frame, $q^{+}=0$, which implies that the form factors are known only for space-like momentum transfer, $q^2=-{q}_\bot^2\leqslant 0$, and the ones in the physical time-like region need  an additional $q^2$ extrapolation.  Following the strategy employed in Refs.~\cite{Jaus:1999zv,Jaus:2002sv,Cheng:2003sm,Shen:2008zzb}, one can parameterize the form factors as functions of $q^2$ by using dipole model in the space-like region and then extend the them to the whole physical region $0\leq q^2\leq (m_{B^*}-m_{V})^2$.   The form factors in the dipole model have the form
 \begin{eqnarray}
F(q^2)=\frac{F(0)}{1-a\,\frac{q^2}{m^2_{B^*}}+b\,(\frac{q^2}{m^2_{B^*}})^2}\,,
\end{eqnarray}
where $F$ denotes $A_{1-4}$  and $V_{1-6}$. Using the inputs given in Table~\ref{tab:LFinput}, we then present our theoretical prediction for the form factors  of $\bar{B}^* \to D^*$, $\bar{B}^*_s \to D^*_s$ and $\bar{B}^*_c \to J/\psi$ transitions in Table \ref{tab:formfactor}. Their $q^2$-dependences are  shown in Fig.~\ref{fig:FFBstar2V}.



\begin{table}[t]
\caption{The numerical results of form factors for $\bar{B}^* \to D^*$, $\bar{B}^*_s \to D^*_s$ and $\bar{B}^*_c \to J/\psi$ transitions within the CLFQM. The uncertainties are caused by the Gaussian parameters listed in Table~\ref{tab:LFinput}. }
\vspace{-0.1cm}\footnotesize
\begin{center}\setlength{\tabcolsep}{3pt}
\begin{tabular}{ccccccccccc}
\hline\hline
&\multicolumn{10}{c}{$\bar{B}^* \to D^*$} \\\hline
             &$A_1$       &$A_2$    &$A_3$    &$A_4$  &$V_1$       &$V_2$    &$V_3$    &$V_4$   &$V_5$    &$V_6$\\\hline
$F(0)$ &$0.66^{+0.01}_{-0.01}$&$0.36^{+0.00}_{-0.00}$&$0.07^{+0.00}_{-0.00}$&$0.08^{+0.00}_{-0.00}$&$0.67^{+0.01}_{-0.01}$&$0.36^{+0.00}_{-0.00}$&$0.13^{+0.00}_{-0.00}$&$0.00^{+0.00}_{-0.00}$&$1.17^{+0.01}_{-0.01}$&$0.48^{+0.01}_{-0.01}$\\
    a
&$1.31^{+0.02}_{-0.02}$&$1.32^{+0.02}_{-0.02}$&$1.79^{+0.02}_{-0.02}$&$1.81^{+0.02}_{-0.02}$&$1.30^{+0.02}_{-0.02}$&$1.32^{+0.02}_{-0.02}$&$1.72^{+0.02}_{-0.02}$&$-0.09^{+0.45}_{-0.40}$&$1.30^{+0.02}_{-0.02}$&$1.29^{+0.02}_{-0.02}$\\
    b
&$0.42^{+0.02}_{-0.02}$&$0.42^{+0.02}_{-0.02}$&$1.10^{+0.03}_{-0.03}$&$1.15^{+0.04}_{-0.04}$&$0.43^{+0.02}_{-0.02}$&$0.42^{+0.02}_{-0.02}$&$1.01^{+0.03}_{-0.04}$&$1.27^{+0.38}_{-0.28}$&$0.41^{+0.02}_{-0.02}$&$0.40^{+0.02}_{-0.02}$\\\hline
&\multicolumn{10}{c}{$\bar{B}^*_s \to D^*_s$ } \\\hline
$F(0)$
    &$0.65^{+0.01}_{-0.01}$&$0.38^{+0.01}_{-0.01}$&$0.10^{+0.00}_{-0.00}$&$0.09^{+0.00}_{-0.00}$&$0.66^{+0.01}_{-0.01}$&$0.38^{+0.01}_{-0.01}$&$0.15^{+0.00}_{-0.00}$&$-0.02^{+0.00}_{-0.00}$&$1.19^{+0.02}_{-0.02}$&$0.53^{+0.01}_{-0.01}$\\
    a
    &$1.42^{+0.03}_{-0.04}$&$1.47^{+0.03}_{-0.03}$&$1.89^{+0.03}_{-0.03}$&$1.88^{+0.02}_{-0.03}$&$1.43^{+0.03}_{-0.04}$&$1.48^{+0.03}_{-0.03}$&$1.79^{+0.03}_{-0.03}$&$2.22^{+0.04}_{-0.03}$&$1.41^{+0.03}_{-0.03}$&$1.35^{+0.04}_{-0.04}$\\
    b &$0.64^{+0.04}_{-0.05}$&$0.67^{+0.04}_{-0.04}$&$1.33^{+0.05}_{-0.06}$&$1.36^{+0.09}_{-0.07}$&$0.64^{+0.04}_{-0.05}$&$0.67^{+0.04}_{-0.05}$&$1.20^{+0.06}_{-0.06}$&$1.92^{+0.08}_{-0.12}$&$0.61^{+0.04}_{-0.05}$&$0.56^{+0.04}_{-0.05}$\\\hline
    &\multicolumn{10}{c}{$\bar{B}^*_c \to J/\psi$ } \\\hline
$F(0)$
    &$0.55^{+0.01}_{-0.01}$&$0.35^{+0.00}_{-0.00}$&$0.14^{+0.00}_{-0.00}$&$0.15^{+0.01}_{-0.01}$&$0.57^{+0.01}_{-0.01}$&$0.35^{+0.00}_{-0.00}$&$0.21^{+0.00}_{-0.01}$&$-0.01^{+0.01}_{-0.01}$&$1.19^{+0.02}_{-0.02}$&$0.64^{+0.01}_{-0.01}$\\
    a     &$2.48^{+0.07}_{-0.07}$&$2.65^{+0.08}_{-0.08}$&$2.88^{+0.09}_{-0.09}$&$2.88^{+0.08}_{-0.08}$&$2.48^{+0.07}_{-0.07}$&$2.56^{+0.08}_{-0.08}$&$2.75^{+0.08}_{-0.09}$&$3.58^{+0.17}_{-0.12}$&$2.42^{+0.07}_{-0.07}$&$2.32^{+0.06}_{-0.06}$\\
    b
    &$2.71^{+0.20}_{-0.22}$&$2.87^{+0.23}_{-0.26}$&$3.88^{+0.31}_{-0.34}$&$3.90^{+0.30}_{-0.33}$&$2.73^{+0.20}_{-0.22}$&$2.88^{+0.23}_{-0.26}$&$3.51^{+0.29}_{-0.32}$&$6.37^{+0.23}_{-0.13}$&$2.54^{+0.20}_{-0.22}$&$2.33^{+0.17}_{-0.19}$\\
\hline\hline
\end{tabular}
\end{center}
\label{tab:formfactor}
\end{table}

\begin{figure}[ht]
\caption{The $q^2$ dependences of differential decay rates $d\Gamma/dq^2$ (solid lines) and $d\Gamma^L/dq^2$ (dashed lines).}
\begin{center}
\subfigure[]{\includegraphics[scale=0.4]{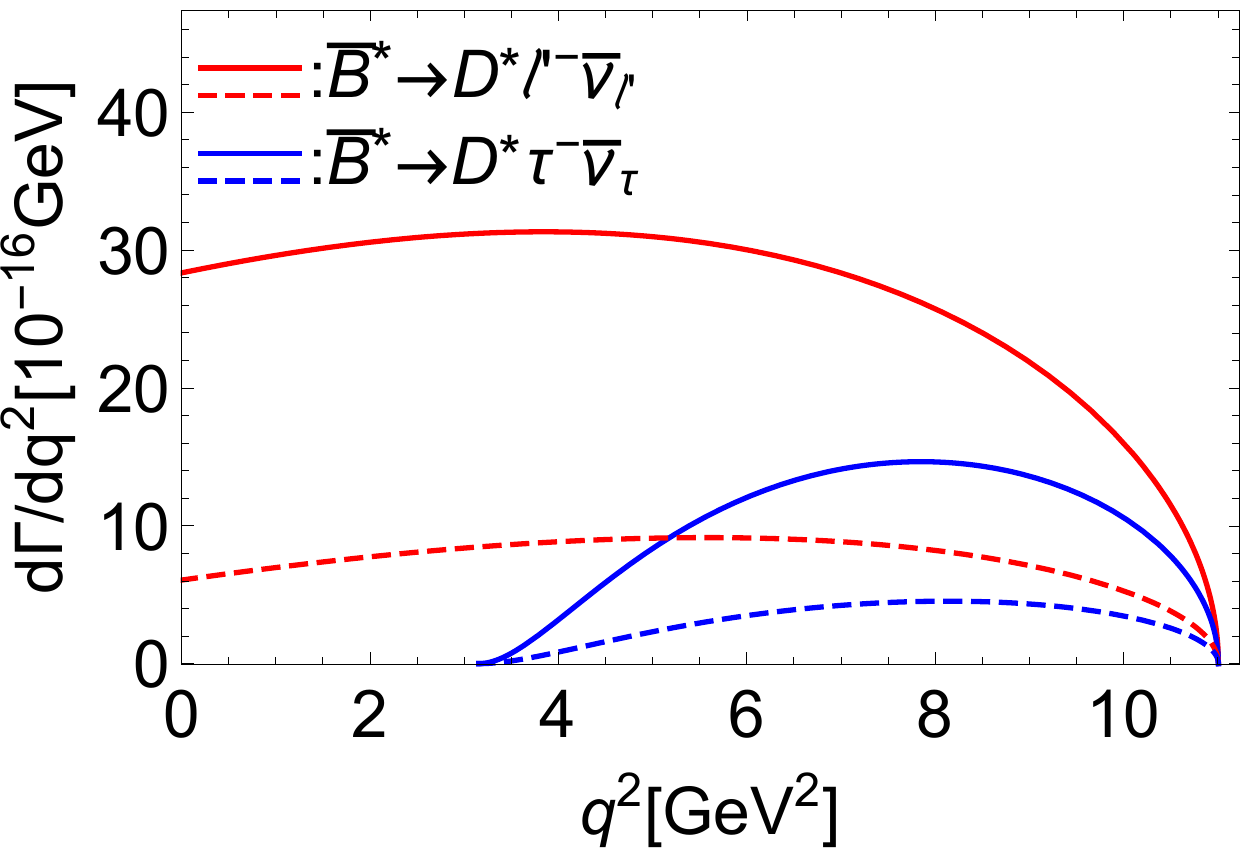}}\qquad
\subfigure[]{\includegraphics[scale=0.4]{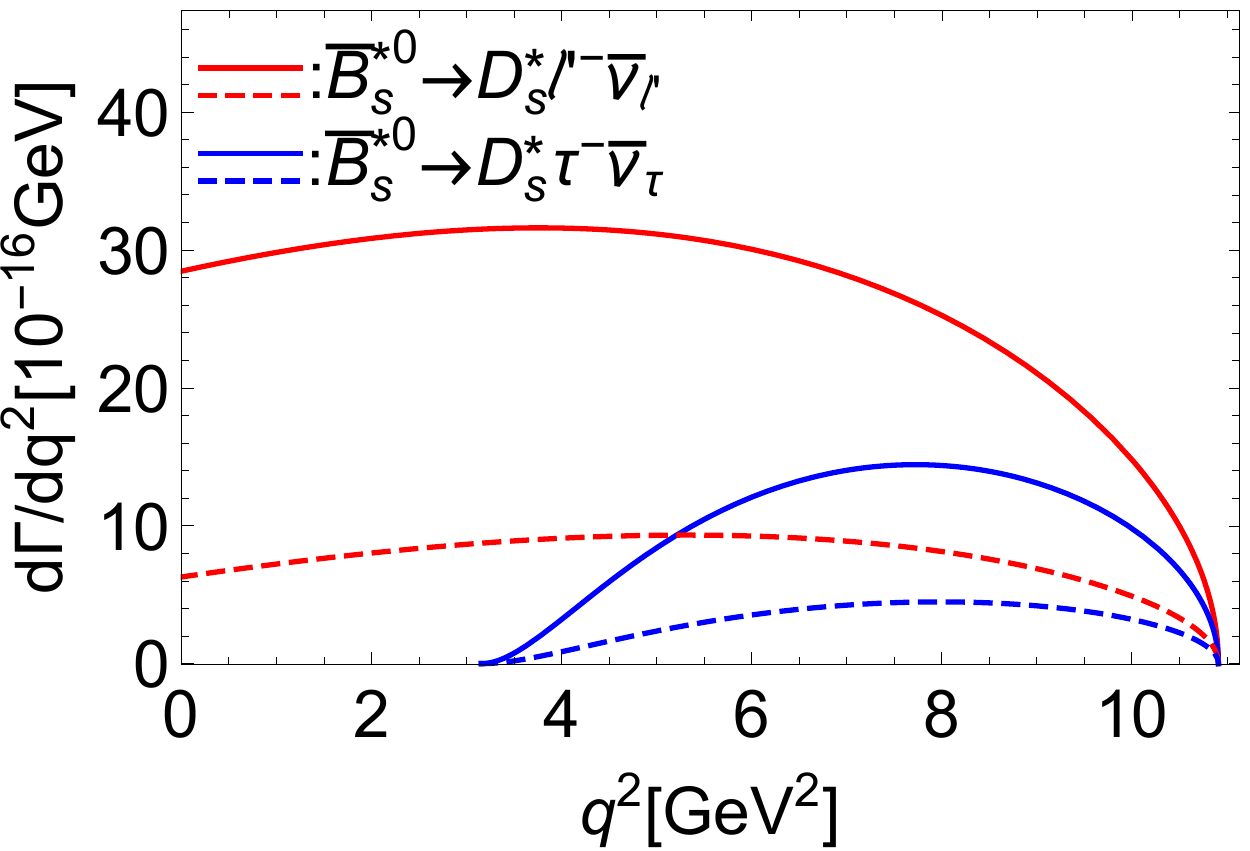}}\qquad
\subfigure[]{\includegraphics[scale=0.4]{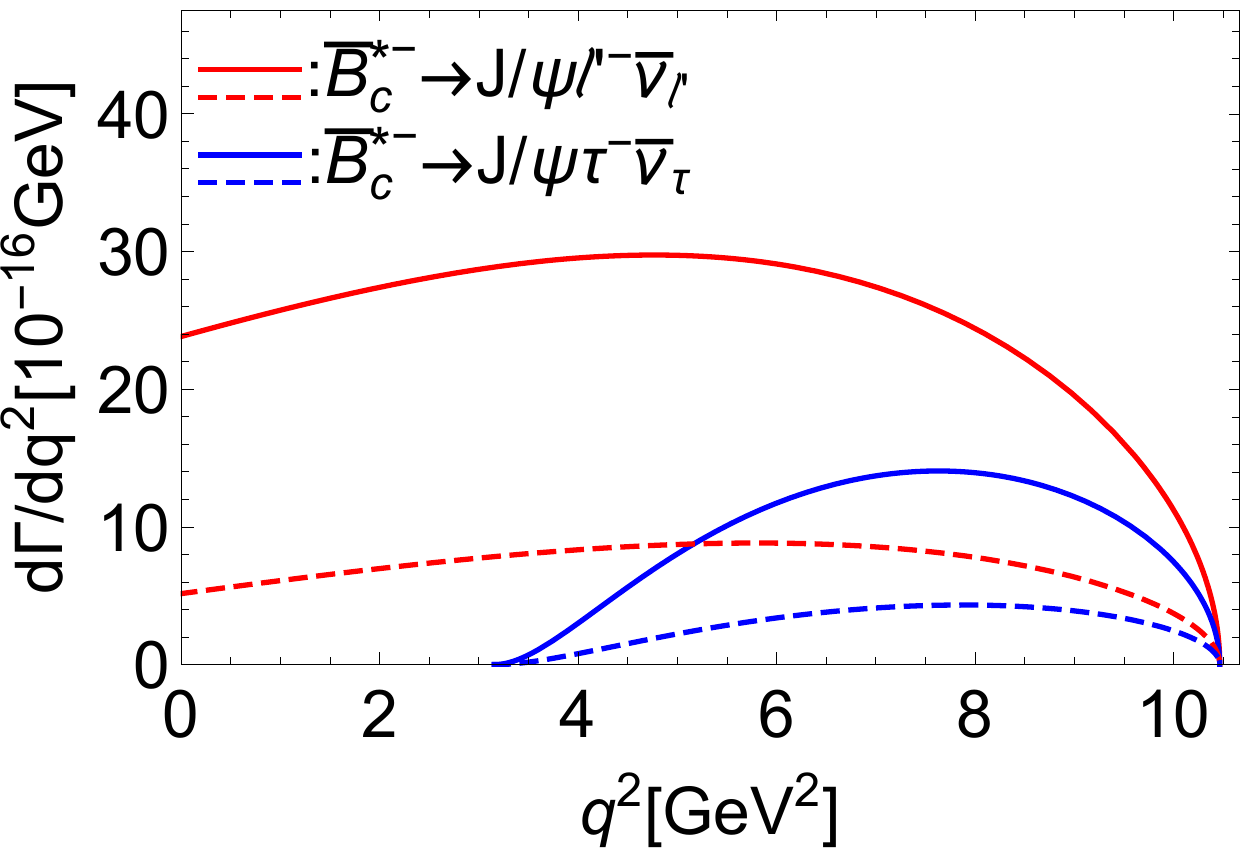}}\\
\end{center}
\label{fig:dGBstar}
\end{figure}

\begin{figure}[ht]
\caption{The $q^2$-dependences of   $R^{*(L)}_{V}(q^2)$, $A^{*V}_{\theta}(q^2)$ and $A^{*V}_{\theta}(q^2)$ .}
\begin{center}
\subfigure[]{\includegraphics[scale=0.45]{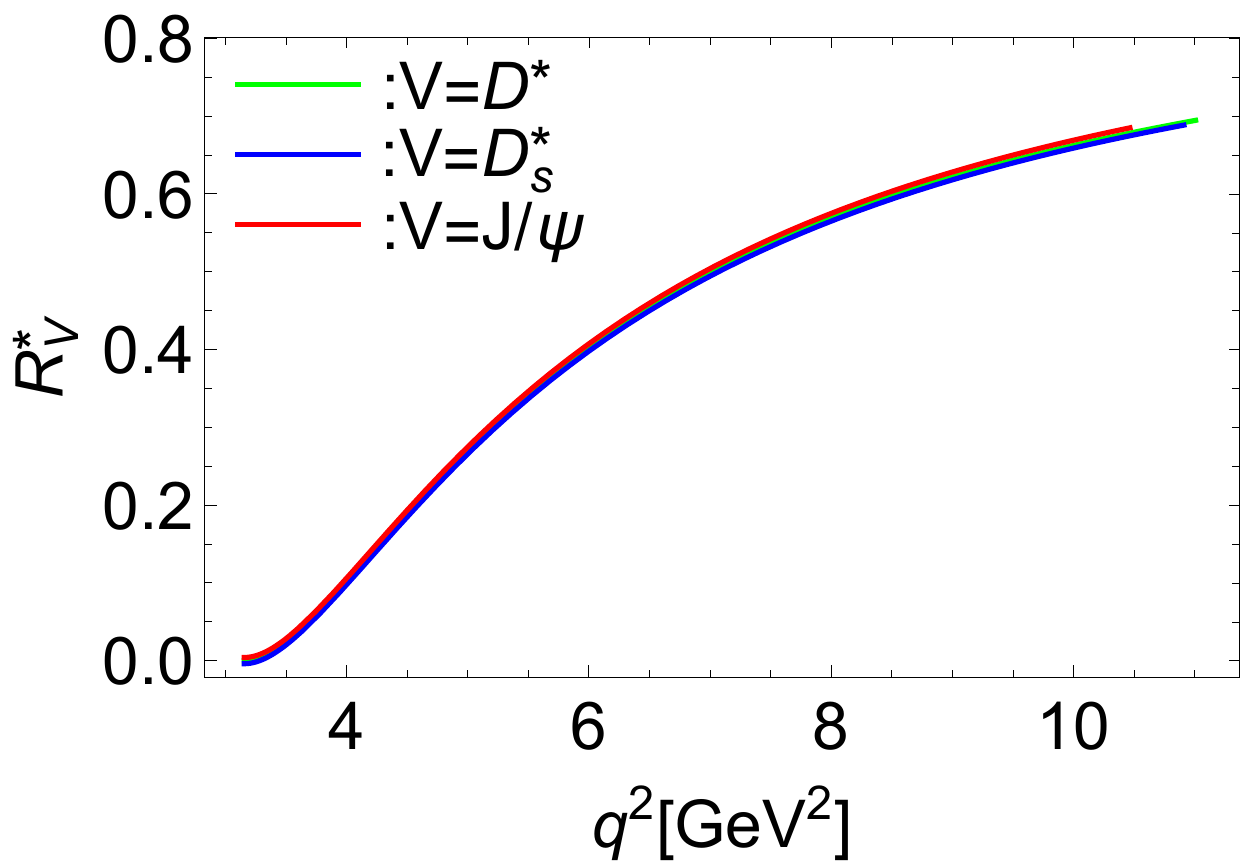}}\qquad
\subfigure[]{\includegraphics[scale=0.45]{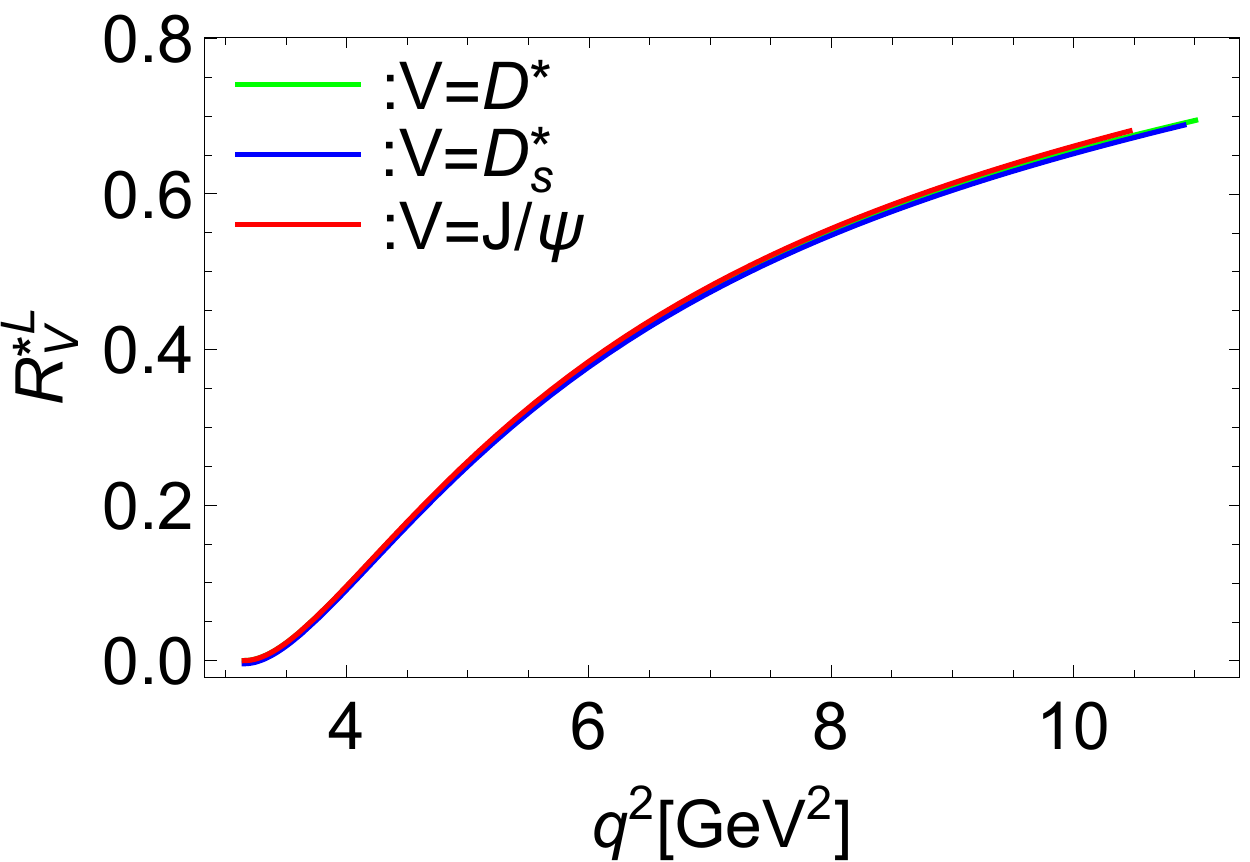}}\\
\subfigure[]{\includegraphics[scale=0.45]{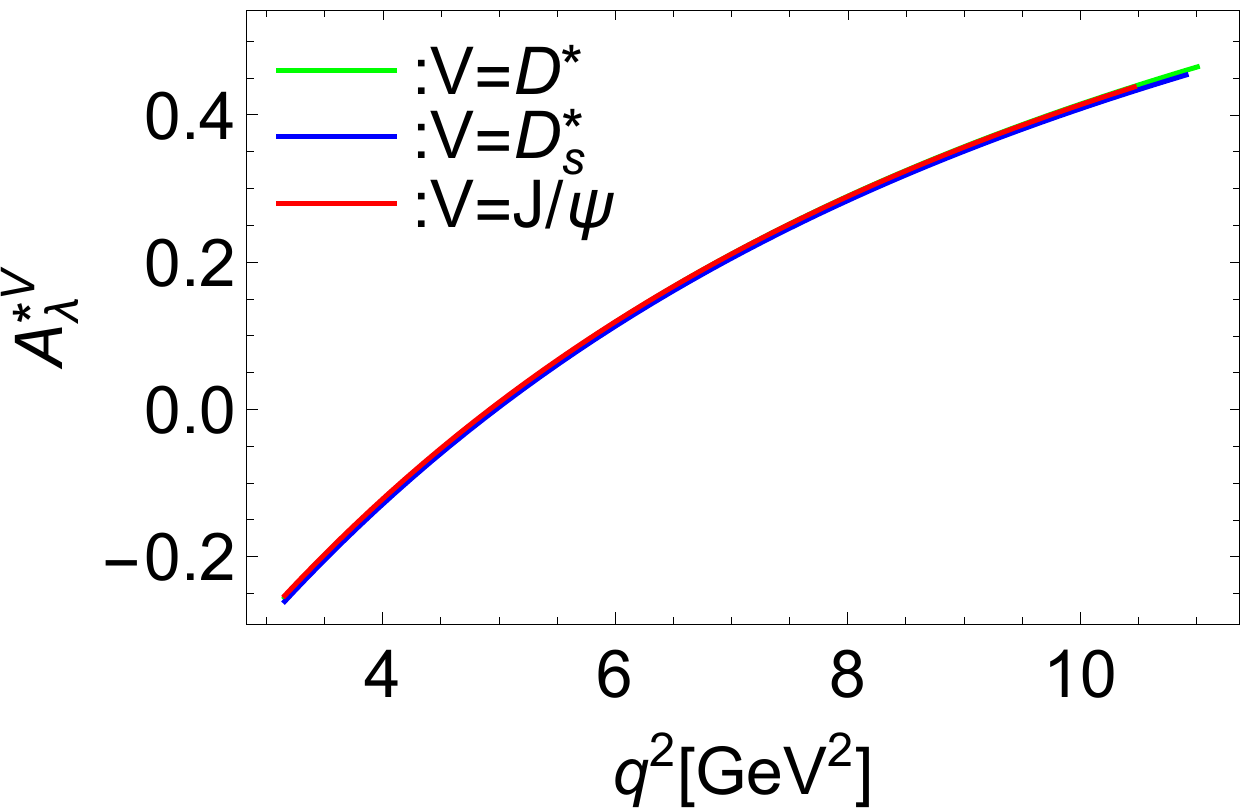}}\qquad
\subfigure[]{\includegraphics[scale=0.45]{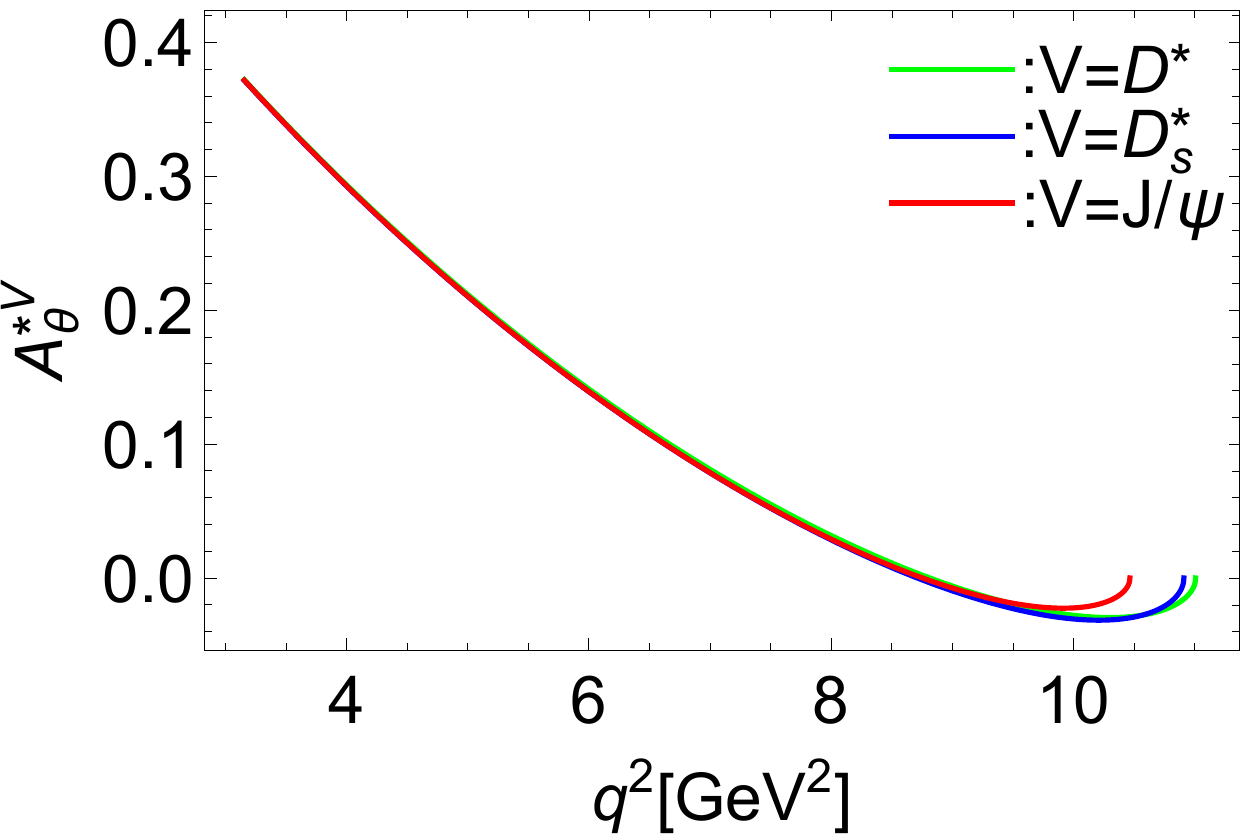}}
\end{center}
\label{fig:dsdobserv}
\end{figure}

Using the formulas given in the last section and inputs given above, we then present our numerical results for the $q^2$-integrated observables of $ \bar{B}^*\to V \ell^- \bar{\nu}_\ell$  decays in Tables~\ref{tab:Bstar2V} and \ref{tab:Obs}. For the branching  fractions, the three errors in  Table~\ref{tab:Bstar2V}  are caused by the uncertainties of form factors, $V_{cb}$ and $\Gamma_{\rm tot}(B^{*})$, respectively. For the other  observables listed in Table~\ref{tab:Obs},  the theoretical uncertainties are  caused only by the form factors. Besides,  the $q^2$-dependence of differential decay rates $d\Gamma^{(L)}/q^2$ and $A_{\lambda\,, \theta}^{*V}$, $R^{*(L)}_V$ are  shown in Figs.~\ref{fig:dGBstar} and \ref{fig:dsdobserv}. The following are some analyses and discussions:

\begin{table}[t]
\caption{The SM predictions for the  branching fractions of  $ \bar{B}^*\to V \ell^- \bar{\nu}_\ell$ decays.}
\begin{center}
\begin{tabular}{lccc}
\hline\hline
Decay mode                                   &This Work& BS method~\cite{Wang:2018dtb}                                 & HQS~\cite{Dai:2018vzz}\\\hline
     $\bar{B}^{*-} \to D^{*0} \ell'^- \bar{\nu}_{\ell'}$
  &$8.42$$^{+0.23}_{-0.24}$$^{+0.11}_{-0.24}$$^{+0.45}_{-0.42}$$\times10^{-8}$&$1.26\times10^{-7}$&$6.41\times10^{-8}$\\
  $\bar{B}^{*-} \to D^{*0} \tau^- \bar{\nu}_{\tau}$
  &$2.26$$^{+0.08}_{-0.08}$$^{+0.03}_{-0.06}$$^{+0.12}_{-0.11}$$\times10^{-8}$&$2.74\times10^{-8}$&$1.29\times10^{-8}$\\
     $\bar{B}^{*0} \to D^{*+} \ell'^- \bar{\nu}_{\ell'}$
  &$2.51$$^{+0.08}_{-0.07}$$^{+0.03}_{-0.07}$$^{+0.13}_{-0.12}$$\times10^{-7}$&$-$&$1.92\times10^{-7}$\\
  $\bar{B}^{*0} \to D^{*+} \tau^- \bar{\nu}_{\tau}$
  &$6.73$$^{+0.24}_{-0.25}$$^{+0.09}_{-0.19}$$^{+0.35}_{-0.32}$$\times10^{-8}$&$-$&$3.88\times10^{-8}$\\
     $\bar{B}^{*0}_s \to D^{*+}_s \ell'^- \bar{\nu}_{\ell'}$
  &$3.46$$^{+0.17}_{-0.17}$$^{+0.05}_{-0.10}$$^{+0.41}_{-0.41}$$\times10^{-7}$&$4.63\times10^{-7}$&$2.53\times10^{-7}$\\
  $\bar{B}^{*0}_s \to D^{*+}_s \tau^- \bar{\nu}_{\tau}$
  &$9.10$$^{+0.60}_{-0.59}$$^{+0.12}_{-0.26}$$^{+1.07}_{-1.07}$$\times10^{-8}$&$1.05\times10^{-7}$&$5.05\times10^{-8}$\\
 $\bar{B}^{*-}_c \to J/\psi \ell'^- \bar{\nu}_{\ell'}$
  &$5.44$$^{+0.33}_{-0.33}$$^{+0.07}_{-0.16}$$^{+4.06}_{-2.00}$$\times10^{-7}$&$5.37\times10^{-7}$&$2.91\times10^{-7}$\\
  $\bar{B}^{*-}_c \to J/\psi \tau^- \bar{\nu}_{\tau}$
  &$1.43$$^{+0.13}_{-0.12}$$^{+0.02}_{-0.04}$$^{+1.07}_{-0.52}$$\times10^{-7}$&$1.49\times10^{-7}$&$5.65\times10^{-8}$\\
  \hline\hline
\end{tabular}
\end{center}
\label{tab:Bstar2V}
\end{table}

\begin{table}[t]
\caption{Predictions for $q^2$-integrated observables  $A_{\lambda\,,\theta}^{*V}$~($\ell=\tau$) , $R^{*(L)}_V$ and $F^{*V}_{L}$.}
\begin{center}
\begin{tabular}{lc lc lc}
\hline\hline
Obs.                                 &Prediction                               &Obs.                            &Prediction                                    &Obs.                          &Prediction   \\\hline
$A^{*D^*}_{\lambda}$     &$0.237^{+0.017}_{-0.016}$    &$A^{*D^*_s}_{\lambda}$     &$0.231^{+0.029}_{-0.030}$         &$A^{*J/\psi}_{\lambda}$ &$0.214^{+0.043}_{-0.040}$\\
$A^{*D^*}_{\theta}$      &$0.070^{+0.007}_{-0.006}$       &$A^{*D^*_s}_{\theta}$     &$0.071^{+0.011}_{-0.011}$           &$A^{*J/\psi}_{\theta}$      &$0.078^{+0.014}_{-0.013}$\\
$R^*_{D^*}$                 &$0.269^{+0.003}_{-0.003}$       &$R^*_{D^*_s}$                 &$0.263^{+0.005}_{-0.005}$           &$R^*_{J/\psi}$                 &$0.262^{+0.009}_{-0.007}$\\
$R^{*L}_{D^*}$            &$0.285^{+0.004}_{-0.003}$        &$R^{*L}_{D^*_s}$            &$0.277^{+0.006}_{-0.007}$            &$R^{*L}_{J/\psi}$             &$0.278^{+0.009}_{-0.009}$\\
$F^{*D^*}_{L}$            &$0.304^{+0.003}_{-0.004}$        &$F^{*D^*_s}_{L}$              &$0.306^{+0.005}_{-0.006}$           &$F^{*J/\psi}_{L}$              &$0.303^{+0.006}_{-0.007}$\\\hline
\hline
\end{tabular}
\end{center}
\label{tab:Obs}
\end{table}
\begin{enumerate}
\item[(1)] From Table \ref{tab:Bstar2V}, one can find a clear relation ${\cal B}(\bar{B}^{*-} \to D^{*0} \ell^- \bar{\nu}_{\ell})$ $:$ ${\cal B}(\bar{B}^{*0} \to D^{*+} \ell^- \bar{\nu}_{\ell})$ $:$ ${\cal B}(\bar{B}^{*0}_{s} \to D^{*+}_{s} \ell^- \bar{\nu}_{\ell})$$:$ ${\cal B}(\bar{B}^{*-}_{c} \to J/\psi \ell^- \bar{\nu}_{\ell})\approx 1:3:4:6$, which is caused mainly by their total decay widths $\Gamma_{\rm tot}(B^*)$ illustrated by  Eqs.~\eqref{eq:GtotBu}, \eqref{eq:GtotBd}, \eqref{eq:GtotBs} and \eqref{eq:GtotBc}.

In Table~\ref{tab:Bstar2V}, the previous predictions based on the Bethe-Salpeter~(BS) method~\cite{Wang:2018dtb}  and  the assumption of heavy quark symmetry~(HQS)~\cite{Dai:2018vzz} are also listed for comparison.   It can be found that the results based on the BS method and the assumption of HQS are  a little bit  smaller and larger respectively than our results, but they are also in agreement  in the order of magnitude. These $b\to c \ell^- \bar{\nu}_{\ell}$ induced $B^*$ weak decays have the  branching fractions of the order $\mathcal{O}(10^{-8}-10^{-7})>10^{-9}$, and therefore are  in the scope of Belle-II or LHC experiments. In addtion, due to the fact that $V_{ub}/V_{cb}\approx0.088$, the $b\to u \ell^- \bar{\nu}_{\ell}$ induced $B^*$ weak decays should have much smaller branching fractions, which are at the level of $\mathcal{O}(10^{-10}-10^{-9})$, and thus are hard to be observed in the near future.

\item[(2)]
Deviations from the SM predictions in $\bar{B} \to D^{(*)} \ell \bar{\nu}_{\ell}$ decay modes have been observed
by the BaBar~\cite{Lees:2012xj,Lees:2013uzd}, Belle~\cite{Huschle:2015rga,Sato:2016svk,Abdesselam:2016xqt} and LHCb~\cite{Aaij:2015yra,Aaij:2017uff} collaborations in the ratios $R_{D^{(*)}}\equiv \frac{\mathcal{B}(\bar{B}\to D^{(*)}\tau^- \bar{\nu}_\tau)}{\mathcal{B}(\bar{B}\to D^{(*)} \ell^{\prime-} \bar{\nu}_{\ell^{\prime}})}$~$(\ell^{\prime}=\mu\,,e)$.  The combination of these measurements performed by the Heavy Flavour Averaging Group (HFLAV)~\cite{Amhis:2016xyh} reads
\begin{equation}
R_{D}^{\rm HFLAV}=0.407\pm0.039\pm0.024\,,\qquad R_{D^{*}}^{\rm HFLAV}=0.306\pm0.013\pm0.007\,,
\end{equation}
which show tensions of about $2$ and $4\sigma$, respectively, with the SM predictions~\cite{Murgui:2019czp}. Very recent measurement of $R_{D^{*}}$ by Belle~\cite{Abdesselam:2019dgh} results in values more compatible with the SM and yield a downward shift in the average. However, even though such measurement is included in the global average, the deviation is still larger than $3\sigma$~\cite{Murgui:2019czp}. If this ``$R_{D^{*}}$ anomaly"  is the truth, it possibly  exists also in the $b\to c$ induced $\bar{B}^*\to V \ell \bar{\nu}_{\ell}$ decays, which therefore can provide another useful test on the  lepton flavor  universality and the various method based on the SM and NP for resolving ``$R_{D^{*}}$ anomaly".  Our numerical results for $R^{*(L)}_{V}$ are summarized in Table~\ref{tab:Obs}, and the $q^2$-spectra of $R^{*(L)}_{V}$ are shown in  Fig.~\ref{fig:dsdobserv}. It can be found that
\begin{equation}
R^{*(L)}_{D^*}\simeq R^{*(L)}_{D^*_s}\simeq R^{*(L)}_{J/\psi}
\end{equation}
 within theoretical uncertainties. Moreover, their  $q^2$-spectra almost overlap with each other as shown in Figs.~\ref{fig:dsdobserv}~(a) and (b).
Using the results summarized  in Table \ref{tab:Bstar2V}, we can also obtain the predictions based on the BS method and HQS,
\begin{eqnarray}
 &&R^*_{D^*}=0.217 \,,~~R^*_{D^*_s}=0.227 \,,~~R^*_{J/\psi}=0.277\,,\qquad\text{BS method} \\
  &&R^*_{D^*}=0.202 \,,~~R^*_{D^*_s}=0.200 \,,~~R^*_{J/\psi}=0.194\,.\qquad\text{HQS}
  \end{eqnarray}
\tb{It can be found that these results are different from our predictions more or less because different models and parameterizations are used for evaluating form factors, which has been observed in the case of $R_{D^*}$~\cite{Jaiswal:2017rve}}.  Future measurement will make a judgement on these results.

\item[(3)] Besides, the  lepton spin asymmetry and the forward-backward asymmetry are also important observables for testing the SM and NP scenarios, for instance, two-Higgs-doublet models, R-parity violating supersymmetry models and so on~\cite{Celis:2012dk,Celis:2016azn,Li:2017jjs,Zhu:2016xdg,Wei:2018vmk,Hu:2018lmk}, because their theoretical uncertainties can be well controlled and the zero-crossing points of their $q^2$-spectra are sensitive to the NP effects~\cite{Celis:2012dk}. Our numerical results for $q^2$-integrated  $A^{*V}_{\lambda}$ and $A^{*V}_{\theta}$ are collected in Table~\ref{tab:Obs}, and the $q^2$ dependences of  $A^{*V}_{\lambda}(q^2)$ and $A^{*V}_{\theta}$ are shown by Figs.~\ref{fig:dsdobserv}~(c) and (d). One can easily find that $A^{*D^*}_{\lambda,\theta}\simeq A^{*D^*_s}_{\lambda,\theta} \simeq A^{*J/\psi}_{\lambda,\theta}$. Moreover,  the $q^2$-spectra of  $A^{*V}_{\lambda}$ are  almost overlap with each other as shown in Fig.~\ref{fig:dsdobserv}~(c), and the case of $A^{*V}_{\theta}$ is similar except that  the $q^2$-spectrum of $A^{*J/\psi}_{\theta}$ deviates from the ones of $A^{*D^*}_{\theta}$ and $A^{*D^*_s}_{\theta}$ at large $q^2$. In addition, $A^{*V}_{\lambda}$ crosses the zero point at $q^2\approx 5{\rm GeV}^2$, however $A^{*V}_{\theta}$ does not have the zero point in all $q^2$ region.

\item[(4)] The $D^*$ longitudinal polarization fraction in semileptonic $B^0 \to D^{\ast -} \tau^+ \nu_{\tau}$ decay, defined as $F^{D^*}_L$=$\Gamma_{\lambda_{D^*}=0}(B^0 \to D^{\ast -} \tau^+ \nu_{\tau})/\Gamma(B^0 \to D^{\ast -} \tau^+ \nu_{\tau})$, has been measured by Belle experiment with $F^{D^*}_L=0.60\pm0.08 (\text{stat.})\pm0.04 (\text{syst.})$~\cite{Abdesselam:2019wbt}, which deviates from the SM prediction  $(F^{D^*}_L)_{\text{SM}}= 0.457\pm0.010$~\cite{Alok:2016qyh} by $1.6\sigma$. Similarly, we can define the longitudinal polarization fraction
\begin{eqnarray}
F^{\ast V}_L=\frac{\Gamma_{\lambda_{D^*}=0}(\bar{B}^{\ast} \to V \tau^- \bar{\nu}_{\tau})}{\Gamma(\bar{B}^{\ast} \to V \tau^- \bar{\nu}_{\tau})}
  \end{eqnarray}
for $\bar{B}^{\ast} \to V \tau^- \bar{\nu}_{\tau}$ decay modes.  From the numerical results given in the last row of Table~\ref{tab:Obs}, one can easily find that
 \begin{eqnarray}
F^{\ast D^*}_L\simeq F^{\ast D_s^*}_L \simeq F^{\ast J/\psi}_L \simeq 30\%,
  \end{eqnarray}
which implies that $\bar{B}^{\ast} \to V \tau^- \bar{\nu}_{\tau}$ decay is dominated by the transverse polarization. It is obviously different from the corresponding $\bar{B} \to V \tau^- \bar{\nu}_{\tau}$ decay mode, which is dominated by the longitudinal polarization state.

 \end{enumerate}

\section{Summary}
In this paper, motivated by abundant $B^*$ data samples  at high-luminosity heavy-flavor experiments in the future, we have studied  the  $b\to c$ induced $\bar{B}^*_{u,d,s,c} \to V \ell^- \bar{\nu}_\ell$ ($V=D^*_{u,d}\,,D^*_s\,,J/\psi$ and $\ell=e\,,\mu\,,\tau$)  decays within the  SM. The helicity amplitudes are investigated in detail, and the form factors of $\bar{B}^*\to V $ transitions are computed within the covariant  light-front quark model. After that we present our predictions for the observables including branching fraction (decay width), leptonic spin asymmetry, forward-backward asymmetry, ratio $R^{\ast(L)}_{V}$ and longitudinal polarization fraction in Tables~\ref{tab:Bstar2V}, \ref{tab:Obs}  and Figs.~\ref{fig:dGBstar},~\ref{fig:dsdobserv}. It is found that all these semileptonic $B^*$ decays have relatively large branching fractions of ${\cal O}(10^{-8})$$\sim$${\cal O}(10^{-7})$, in which $\mathcal{B}(\bar{B}^*_{c} \to J/\psi \ell'^- \bar{\nu}_{\ell'})\sim 5\times 10^{-7}$ is the largest one, and are hopeful to be observed at running LHC and SuperKEKB/Belle-II experiments; in addition, for the  $\bar{B}^{\ast} \to V \tau^- \bar{\nu}_{\tau}$ decay, the  longitudinal polarization state of $V$ meson presents only about $30\%$ contribution to the integrated decay width, which is obviously different from the corresponding $\bar{B} \to V \tau^- \bar{\nu}_{\tau}$ decay. All of  results and findings in this paper are waiting for the experimental test in the future. 

\section*{Data Availability Statement} 
This manuscript has no associated data or the data will not be deposited. This is a theoretical research work, no additional data are associated with this work.

\section*{Acknowledgements}
This work is supported by the National Natural Science Foundation of China (Grant Nos. 11875122 and 11475055) and the Program for Innovative Research Team in University of Henan Province (Grant No.19IRTSTHN018).



\begin{thebibliography}{99}

\bibitem{Amhis:2016xyh}
  Y.~Amhis {\it et al.} [HFLAV Collaboration],
  Eur.\ Phys.\ J.\ C {\bf 77} (2017) no.12,  895.
  
\bibitem{DescotesGenon:2012zf}
 \tb{ S.~Descotes-Genon, J.~Matias, M.~Ramon and J.~Virto,
  JHEP {\bf 1301} (2013) 048.}

\bibitem{Aaij:2015oid}
  \tb{ R.~Aaij {\it et al.} [LHCb Collaboration],
  JHEP {\bf 1602} (2016) 104.
  }
 
 \bibitem{Khachatryan:2015isa}
\tb{  V.~Khachatryan {\it et al.} [CMS Collaboration],
  Phys.\ Lett.\ B {\bf 753} (2016) 424.}

\bibitem{Aaboud:2018krd}
 \tb{ M.~Aaboud {\it et al.} [ATLAS Collaboration],
  JHEP {\bf 1810} (2018) 047.}
  
\bibitem{Wehle:2016yoi}
 \tb{ S.~Wehle {\it et al.} [Belle Collaboration],
  Phys.\ Rev.\ Lett.\  {\bf 118} (2017) no.11,  111801.}

\bibitem{Aaij:2015esa}
  \tb{ R.~Aaij {\it et al.} [LHCb Collaboration],
  JHEP {\bf 1509} (2015) 179}

\bibitem{Aaij:2013aln}
\tb{  R.~Aaij {\it et al.} [LHCb Collaboration],
  JHEP {\bf 1307} (2013) 084.}

\bibitem{Buras:2003dj}
\tb{  A.~J.~Buras, R.~Fleischer, S.~Recksiegel and F.~Schwab,
  Phys.\ Rev.\ Lett.\  {\bf 92} (2004) 101804.}

\bibitem{Buras:2004ub}
 \tb{ A.~J.~Buras, R.~Fleischer, S.~Recksiegel and F.~Schwab,
  Nucl.\ Phys.\ B {\bf 697} (2004) 133. }

 \bibitem{Lees:2012xj}
  J.~P.~Lees {\it et al.} [BaBar Collaboration],
  Phys.\ Rev.\ Lett.\  {\bf 109} (2012) 101802.

  \bibitem{Lees:2013uzd}
  J.~P.~Lees {\it et al.} [BaBar Collaboration],
  Phys.\ Rev.\ D {\bf 88} (2013) no. 7, 072012.

  \bibitem{Huschle:2015rga}
  M.~Huschle {\it et al.} [Belle Collaboration],
  Phys.\ Rev.\ D {\bf 92} (2015) no.7,  072014.


\bibitem{Sato:2016svk}
  Y.~Sato {\it et al.} [Belle Collaboration],
  Phys.\ Rev.\ D {\bf 94} (2016) no.7,  072007.

\bibitem{Abdesselam:2016xqt}
  A.~Abdesselam {\it et al.}  \tb{[Belle Collaboration]},
  arXiv:1608.06391 [hep-ex].

\bibitem{Aaij:2015yra}
  R.~Aaij {\it et al.} [LHCb Collaboration],
  Phys.\ Rev.\ Lett.\  {\bf 115} (2015) no.11,  111803
   Addendum: [Phys.\ Rev.\ Lett.\  {\bf 115} (2015) no.15,  159901].

\bibitem{Aaij:2017uff}
  R.~Aaij {\it et al.} [LHCb Collaboration],
  Phys.\ Rev.\ Lett.\  {\bf 120} (2018) no.17,  171802.


  
\bibitem{Jaiswal:2017rve}
\tb{  S.~Jaiswal, S.~Nandi and S.~K.~Patra,
  JHEP {\bf 1712} (2017) 060.}

\bibitem{Jaiswal:2020wer}
\tb{  S.~Jaiswal, S.~Nandi and S.~K.~Patra,
  arXiv:2002.05726 [hep-ph].}
  

\bibitem{Bhattacharya:2014wla}
 \tb{ B.~Bhattacharya, A.~Datta, D.~London and S.~Shivashankara,
  Phys.\ Lett.\ B {\bf 742} (2015) 370.}
  
\bibitem{Bhattacharya:2015ida}
 \tb{ S.~Bhattacharya, S.~Nandi and S.~K.~Patra,
  Phys.\ Rev.\ D {\bf 93} (2016) no.3,  034011.}

\bibitem{Chen:2005gr}
\tb{  C.~H.~Chen and C.~Q.~Geng,
  Phys.\ Rev.\ D {\bf 71} (2005) 077501.}
    
\bibitem{Bhattacharya:2016zcw}
 \tb{ S.~Bhattacharya, S.~Nandi and S.~K.~Patra,
  Phys.\ Rev.\ D {\bf 95} (2017) no.7,  075012.}
  
\bibitem{Bhattacharya:2018kig}
 \tb{ S.~Bhattacharya, S.~Nandi and S.~Kumar Patra,
  Eur.\ Phys.\ J.\ C {\bf 79} (2019) no.3,  268.}

\bibitem{Alok:2017qsi}
 \tb{ A.~K.~Alok, D.~Kumar, J.~Kumar, S.~Kumbhakar and S.~U.~Sankar,
  JHEP {\bf 1809} (2018) 152.}
   
\bibitem{Feruglio:2018fxo}
 \tb{ F.~Feruglio, P.~Paradisi and O.~Sumensari,
  JHEP {\bf 1811} (2018) 191.}

\bibitem{Azatov:2018knx}
\tb{  A.~Azatov, D.~Bardhan, D.~Ghosh, F.~Sgarlata and E.~Venturini,
  JHEP {\bf 1811} (2018) 187.}

\bibitem{Freytsis:2015qca}
 \tb{ M.~Freytsis, Z.~Ligeti and J.~T.~Ruderman,
  Phys.\ Rev.\ D {\bf 92} (2015) no.5,  054018.}

\bibitem{Li:2016vvp}
 \tb{ X.~Q.~Li, Y.~D.~Yang and X.~Zhang,
  JHEP {\bf 1608} (2016) 054.}

\bibitem{Hiller:2016kry}
\tb{  G.~Hiller, D.~Loose and K.~Schönwald,
  JHEP {\bf 1612} (2016) 027.}
 
\bibitem{Blanke:2018sro}
 \tb{ M.~Blanke and A.~Crivellin,
  Phys.\ Rev.\ Lett.\  {\bf 121} (2018) no.1,  011801.}
 
\bibitem{Kim:2015zla}
\tb{  C.~S.~Kim, Y.~W.~Yoon and X.~B.~Yuan,
  JHEP {\bf 1512} (2015) 038.}
  
\bibitem{Crivellin:2015hha}
  \tb{A.~Crivellin, J.~Heeck and P.~Stoffer,
  Phys.\ Rev.\ Lett.\  {\bf 116} (2016) no.8,  081801.}
  
\bibitem{He:2017bft}
  \tb{X.~G.~He and G.~Valencia,
  Phys.\ Lett.\ B {\bf 779} (2018) 52.}
  
\bibitem{Hu:2020yvs}
\tb{  Q.~Y.~Hu, Y.~D.~Yang and M.~D.~Zheng,
  arXiv:2002.09875 [hep-ph].  }

\bibitem{Wang:2019trs}
\tb{  D.~Y.~Wang, Y.~D.~Yang and X.~B.~Yuan,
  Chin.\ Phys.\ C {\bf 43} (2019) no.8,  083103.}

\bibitem{Yan:2019hpm}
 \tb{ H.~Yan, Y.~D.~Yang and X.~B.~Yuan,
  Chin.\ Phys.\ C {\bf 43} (2019) no.8,  083105.}

\bibitem{Hu:2018veh}
  \tb{Q.~Y.~Hu, X.~Q.~Li and Y.~D.~Yang,
  Eur.\ Phys.\ J.\ C {\bf 79} (2019) no.3,  264.}

\bibitem{Li:2018rax}
\tb{  S.~P.~Li, X.~Q.~Li, Y.~D.~Yang and X.~Zhang,
  JHEP {\bf 1809} (2018) 149.}

\bibitem{Altmannshofer:2017poe}
  \tb{W.~Altmannshofer, P.~S.~Bhupal Dev and A.~Soni,
  Phys.\ Rev.\ D {\bf 96} (2017) no.9,  095010.}
             
\bibitem{Cheung:2020sbq}
\tb{  K.~Cheung, Z.~R.~Huang, H.~D.~Li, C.~D.~Lü, Y.~N.~Mao and R.~Y.~Tang,
  arXiv:2002.07272 [hep-ph].}

 \bibitem{Celis:2012dk}
  A.~Celis, M.~Jung, X.~Q.~Li and A.~Pich,
  JHEP {\bf 1301} (2013) 054.

\bibitem{Celis:2016azn}
  A.~Celis, M.~Jung, X.~Q.~Li and A.~Pich,
  Phys.\ Lett.\ B {\bf 771} (2017) 168.

\bibitem{Li:2017jjs}
  X.~Q.~Li,
  Nucl.\ Part.\ Phys.\ Proc.\  {\bf 287-288} (2017) 181.


  \bibitem{Zhu:2016xdg}
  J.~Zhu, H.~M.~Gan, R.~M.~Wang, Y.~Y.~Fan, Q.~Chang and Y.~G.~Xu,
  Phys.\ Rev.\ D {\bf 93} (2016) no.9,  094023.

  \bibitem{Wei:2018vmk}
  J.~Zhu, B.~Wei, J.~H.~Sheng, R.~M.~Wang, Y.~Gao and G.~R.~Lu,
  Nucl.\ Phys.\ B {\bf 934} (2018) 380.

  \bibitem{Hu:2018lmk}
  Q.~Y.~Hu, X.~Q.~Li, Y.~Muramatsu and Y.~D.~Yang,
  Phys.\ Rev.\ D {\bf 99} (2019) no.1,  015008.

    
  
\bibitem{Bifani:2018zmi}
\tb{  S.~Bifani, S.~Descotes-Genon, A.~Romero Vidal and M.~H.~Schune,
  J.\ Phys.\ G {\bf 46} (2019) no.2,  023001.}
  
\bibitem{Li:2018lxi}
 \tb{ Y.~Li and C.~D.~Lü,
  Sci.\ Bull.\  {\bf 63} (2018) 267.}
  


  

  
  
  
  
    
\bibitem{Isgur:1991wq}
  N.~Isgur and M.~B.~Wise,
  Phys.\ Rev.\ Lett.\  {\bf 66} (1991) 1130.

\bibitem{Godfrey:1986wj}
  S.~Godfrey and R.~Kokoski,
  Phys.\ Rev.\ D {\bf 43} (1991) 1679.

\bibitem{Eichten:1993ub}
  E.~J.~Eichten, C.~T.~Hill and C.~Quigg,
  Phys.\ Rev.\ Lett.\  {\bf 71} (1993) 4116.

\bibitem{Ebert:1997nk}
  D.~Ebert, V.~O.~Galkin and R.~N.~Faustov,
  Phys.\ Rev.\ D {\bf 57} (1998) 5663
   [Erratum Phys.\ Rev.\ D {\bf 59} (1998) 019902].
   
\bibitem{Tanabashi:2018oca}
  M.~Tanabashi {\it et al.} [Particle Data Group],
  Phys.\ Rev.\ D {\bf 98} (2018) no.3,  030001.

\bibitem{Abe:2010gxa}
  T.~Abe {\it et al.}  [Belle-II Collaboration], arXiv:1011.0352.
  
\bibitem{Huang:2006mf}
\tb{  G.~S.~Huang {\it et al.}  (CLEO Collaboration), hep-ex/0607080.}


\bibitem{Aaij:2010gn}
  R. Aaij {\it et al.}  (LHCb Collaboration),  Phys.  Lett. B {\bf 694} (2010) 209.

\bibitem{Bediaga:2012py}
  R.~Aaij {\it et al.} [LHCb Collaboration],
  Eur.\ Phys.\ J.\ C {\bf 73} (2013) no.4,  2373.

\bibitem{Aaij:2014jba}
  R.~Aaij {\it et al.} [LHCb Collaboration],
  Int.\ J.\ Mod.\ Phys.\ A {\bf 30} (2015) no.07,  1530022.

\bibitem{Grinstein:2015aua}
  B.~Grinstein and J.~Martin Camalich,
  Phys.\ Rev.\ Lett.\  {\bf 116} (2016) no.14,  141801.

  
\bibitem{Xu:2015eev}
  G.~Z.~Xu, Y.~Qiu, C.~P.~Shen and Y.~J.~Zhang,
  Eur.\ Phys.\ J.\ C {\bf 76} (2016) no.11,  583.

\bibitem{Wang:2012hu}
  Z.~G.~Wang,
  Commun.\ Theor.\ Phys.\  {\bf 61} (2014) 1,  81.

 \bibitem{Zeynali:2014wya}
  K.~Zeynali, V.~Bashiry and F.~Zolfagharpour,
  Eur.\ Phys.\ J.\ A {\bf 50} (2014) 127.

\bibitem{Bashiry:2014qia}
  V.~Bashiry,
  Adv.\ High Energy Phys.\  {\bf 2014} (2014) 503049.

\bibitem{Wang:2018dtb}
  T.~Wang, Y.~Jiang, T.~Zhou, X.~Z.~Tan and G.~L.~Wang,
  J.\ Phys.\ G {\bf 45} (2018) no.11,  115001.

\bibitem{Dai:2018vzz}
  L.~R.~Dai, X.~Zhang and E.~Oset,
  Phys.\ Rev.\ D {\bf 98} (2018) no.3,  036004.

\bibitem{Chang:2016cdi}
  Q.~Chang, J.~Zhu, X.~L.~Wang, J.~F.~Sun and Y.~L.~Yang,
  Nucl.\ Phys.\ B {\bf 909} (2016) 921.

 \bibitem{Chang:2015jla}
  Q.~Chang, P.~P.~Li, X.~H.~Hu and L.~Han,
  Int.\ J.\ Mod.\ Phys.\ A {\bf 30} (2015) no.27,  1550162.

  \bibitem{Chang:2015ead}
  Q.~Chang, X.~Hu, J.~Sun, X.~Wang and Y.~Yang,
  Adv.\ High Energy Phys.\  {\bf 2015} (2015) 767523.
  
\bibitem{Chang:2016eto}
  Q.~Chang, L.~X.~Chen, Y.~Y.~Zhang, J.~F.~Sun and Y.~L.~Yang,
  Eur.\ Phys.\ J.\ C {\bf 76} (2016) no.10,  523.

\bibitem{Sun:2017lup}
  J.~Sun, Y.~Yang, N.~Wang, Q.~Chang and G.~Lu,
  Phys.\ Rev.\ D {\bf 95} (2017) no.7,  074032.

\bibitem{Chang:2018mva}
  Q.~Chang, L.~L.~Chen and S.~Xu,
  J.\ Phys.\ G {\bf 45} (2018) no.7,  075005.

\bibitem{Sun:2017xed}
  J.~Sun, H.~Li, Y.~Yang, N.~Wang, Q.~Chang and G.~Lu,
  J.\ Phys.\ G {\bf 44} (2017) no.7,  075007.

\bibitem{Sun:2017lla}
  J.~Sun, Y.~Yang, N.~Wang, J.~Huang and Q.~Chang,
  Phys.\ Rev.\ D {\bf 95} (2017) no.3,  036024.

\bibitem{Chang:2018sud}
  Q.~Chang, J.~Zhu, N.~Wang and R.~M.~Wang,
  Adv.\ High Energy Phys.\  {\bf 2018} (2018) 7231354

\bibitem{Zhang:2019hth}
  J.~Zhang, Y.~Zhang, Q.~Zeng and R.~Sun,
  Eur.\ Phys.\ J.\ C {\bf 79} (2019) no.2,  164.

\bibitem{Korner:1987kd}
  J.~G.~Korner and G.~A.~Schuler,
  Z.\ Phys.\ C {\bf 38} (1988) 511 [Erratum: Z.\ Phys.\ C {\bf 41} (1989) 690].

\bibitem{Korner:1989qb}
  J.~G.~Korner and G.~A.~Schuler,
  Z.\ Phys.\ C {\bf 46} (1990) 93.

 \bibitem{Wang:2007ys}
  Y.~M.~Wang, H.~Zou, Z.~T.~Wei, X.~Q.~Li and C.~D.~Lu,
  Eur.\ Phys.\ J.\ C {\bf 54} (2008) 107.

\bibitem{Shen:2008zzb}
  Y.~L.~Shen and Y.~M.~Wang,
  Phys.\ Rev.\ D {\bf 78} (2008) 074012.
    
 \bibitem{Fajfer:2012vx}
  S.~Fajfer, J.~F.~Kamenik and I.~Nisandzic,
  Phys.\ Rev.\ D {\bf 85} (2012) 094025.

\bibitem{Kadeer:2005aq}
  A.~Kadeer, J.~G.~Korner and U.~Moosbrugger,
  Eur.\ Phys.\ J.\ C {\bf 59} (2009) 27.

\bibitem{Charles:2004jd}
 J. Charles {\it et al.} (CKMfitter Group), Eur. Phys. J. C {\bf 41} (2005) 1;
 updated results and plots available at: http://ckmfitter.in2p3.fr.

  \bibitem{Choi:2007se}
  H.~M.~Choi,
  Phys.\ Rev.\ D {\bf 75} (2007) 073016.

\bibitem{Verma:2011yw}
  R.~C.~Verma,
  J.\ Phys.\ G {\bf 39} (2012) 025005.

  \bibitem{Chang:2018zjq}
  Q.~Chang, X.~N.~Li, X.~Q.~Li, F.~Su and Y.~D.~Yang,
  Phys.\ Rev.\ D {\bf 98} (2018) no.11,  114018.



 \bibitem{Goity:2000dk}
  J.~L.~Goity and W.~Roberts,
  Phys.\ Rev.\ D {\bf 64} (2001) 094007.

\bibitem{Ebert:2002xz}
  D.~Ebert, R.~N.~Faustov and V.~O.~Galkin,
  Phys.\ Lett.\ B {\bf 537} (2002) 241.

\bibitem{Zhu:1996qy}
  S.~L.~Zhu, W.~Y.~P.~Hwang and Z.~s.~Yang,
  Mod.\ Phys.\ Lett.\ A {\bf 12} (1997) 3027.

\bibitem{Aliev:1995wi}
  T.~M.~Aliev, D.~A.~Demir, E.~Iltan and N.~K.~Pak,
  Phys.\ Rev.\ D {\bf 54} (1996) 857.

\bibitem{Colangelo:1993zq}
  P.~Colangelo, F.~De Fazio and G.~Nardulli,
  Phys.\ Lett.\ B {\bf 316} (1993) 555.

\bibitem{Cheung:2014cka}
  C.~Y.~Cheung and C.~W.~Hwang,
  JHEP {\bf 1404} (2014) 177.


\bibitem{Jaus:1999zv}
  W.~Jaus,
  Phys.\ Rev.\ D {\bf 60}  (1999)  054026.

\bibitem{Jaus:2002sv}
  W.~Jaus,
  Phys.\ Rev.\ D {\bf 67}  (2003) 094010.

\bibitem{Cheng:2003sm}
  H.~Y.~Cheng, C.~K.~Chua and C.~W.~Hwang,
  Phys.\ Rev.\ D {\bf 69}  (2004) 074025.


\bibitem{Chang:2019xtj}
  Q.~Chang, Y.~Zhang and X.~Li,
\tb{  Chin.\ Phys.\ C {\bf 43} (2019) no.10,  103104.}
  
\bibitem{Murgui:2019czp}
  C.~Murgui, A.~Penuelas, M.~Jung and A.~Pich,
\tb{  JHEP {\bf 1909} (2019) 103.}

  
\bibitem{Abdesselam:2019dgh}
  A.~Abdesselam {\it et al.} [Belle Collaboration],
  arXiv:1904.08794 [hep-ex].
  


 

  \bibitem{Abdesselam:2019wbt}
  A.~Abdesselam {\it et al.} [Belle Collaboration],
  arXiv:1903.03102 [hep-ex].

  \bibitem{Alok:2016qyh}
  A.~K.~Alok, D.~Kumar, S.~Kumbhakar and S.~U.~Sankar,
  Phys.\ Rev.\ D {\bf 95} (2017) no.11,  115038.






\end{thebibliography}
\end{document}